\documentclass[amsmath,amssymb,aps,prd,nofootinbib,twocolumn,superscriptaddress]{revtex4-2}
\usepackage[utf8]{inputenc}
\usepackage{mathrsfs}
\usepackage{bm}
\usepackage{url}
\usepackage{mathtools}
\usepackage{array}
\newcolumntype{P}[1]{>{\centering\arraybackslash}p{#1}}
\newcolumntype{M}[1]{>{\centering\arraybackslash}m{#1}}
\usepackage[caption=false]{subfig}
\usepackage{dcolumn}
\usepackage{graphicx, epsfig}
\usepackage[dvipsnames]{xcolor}
\usepackage{mathrsfs}
\usepackage{bm}
\usepackage{yhmath}
\usepackage[caption=false]{subfig}
\usepackage[normalem]{ulem}
\usepackage{mathtools}
\usepackage{bigints}
\usepackage{float}
\usepackage{afterpage}
\usepackage{threeparttable}
\usepackage{tablefootnote}
\usepackage[colorlinks = true, linkcolor = purple, urlcolor  = blue, citecolor = blue, anchorcolor = blue]{hyperref}
\usepackage{float}
\usepackage{multirow}
\usepackage{comment}

\newcommand{\be}{\begin{equation}}
\newcommand{\ee}{\end{equation}}
\newcommand{\bs}{\begin{split}} 
\newcommand{\bea}{\begin{eqnarray}}
\newcommand{\eea}{\end{eqnarray}}

\begin{document}


\title{Interpreting the diversity of afterglow emission from radio-detected tidal disruption events with instantaneous and delayed outflows}

\author{Yuri Sato}
\email{yuri@astr.tohoku.ac.jp (YS)}
\affiliation{Astronomical Institute, Graduate School of Science, Tohoku University, Sendai 980-8578, Japan}
\author{Mukul Bhattacharya}
\affiliation{Department of Physics, Wisconsin IceCube Particle Astrophysics Center, University of Wisconsin, Madison, WI 53703, USA}
\affiliation{Department of Astronomy, Astrophysics and Space Engineering, Indian Institute of Technology Indore, Simrol, MP 453552, India}
\author{Jose Alonso Carpio}
\affiliation{Department of Physics \& Astronomy; Nevada Center for Astrophysics, University of Nevada, Las Vegas, NV 89154, USA
}
\author{Jewel~Capili}
\affiliation{Department of Physics, California Polytechnic State University, San Luis Obispo, CA, 93401, USA}
\affiliation{Department of Religious Studies, Naropa University, Boulder, CO, 80302}
\author{Kohta Murase}
\affiliation{Department of Physics; Department of Astronomy \& Astrophysics; Center for Multimessenger Astrophysics, Institute for Gravitation and the Cosmos, The Pennsylvania State University, University Park, PA 16802, USA}
\affiliation{Center for Gravitational Physics and Quantum Information, Yukawa Institute for Theoretical Physics, Kyoto University, Kyoto, Kyoto 606-8502, Japan}

\date{\today}

\begin{abstract}
Tidal disruption events (TDEs) occur when a star is gravitationally disrupted by the tidal field of a supermassive black hole during a close encounter.
Radio emission has recently been detected in TDEs and is commonly attributed to synchrotron radiation from both wind and jetted outflows.
However, several TDEs exhibit bright radio flares at late times, which cannot be easily explained if the wind is launched promptly after the stellar disruption.
The radio light curves of TDEs with delayed radio flares are modeled using three scenarios: an instantaneous wind, a delayed wind, and a delayed relativistic jet. We show that the instantaneous wind model struggles to reproduce delayed radio flare events, indicating the necessity of an additional delayed outflow component. In contrast, the delayed wind model provides a consistent explanation for the observed radio phenomenology, successfully reproducing events both with and without delayed radio flares. 
For some delayed radio flare events (e.g., ASASSN-15oi and AT~2019dsg), both the delayed wind and delayed jet models can reproduce the observed radio light curves, 
 making it challenging to distinguish between these scenarios based on radio data alone. 
To break this degeneracy, we present calculations of detailed multiwavelength spectra from radio to x-ray and very-high-energy gamma-ray bands, including both synchrotron and inverse-Compton processes. Within this framework,
the delayed jet model produces x-ray and optical emission detectable at typical TDE distances, in contrast to wind-driven scenarios. Our results highlight how multiwavelength campaigns offer effective means to distinguish between possible outflow mechanisms. We apply our model to AT 2025aarm, which is one of the closest TDEs that has been observed, and show the importance of x-ray observations.
\end{abstract}


\maketitle



\section{Introduction}
A tidal disruption event (TDE) occurs when a star is torn apart by the tidal forces of a supermassive black hole (SMBH) during a close encounter \citep{Hills1975,Rees1988,EK1989}. 
The stellar debris is divided into roughly equal components:
one portion becomes unbound and is ejected into space, while the remainder returns toward the SMBH to form an accretion flow. 
This process can give rise to multiwavelength emission in radio, infrared (IR)/optical/ultraviolet (UV), x-ray and gamma-ray bands.
While the observed thermal emission is commonly interpreted as reprocessed radiation arising from the accreting stellar debris \citep{Loeb1997,Strubbe2009,Lodato2011,Komossa2015,MS2016,Roth2016}, alternative interpretations have also been proposed. These include the collisions of debris streams (e.g., Ref.~\citep{Krolik:2024tvu}), the gravitational contraction of a quasispherical envelope~\citep{Metzger:2022lob}, and late-time emission associated with a spreading accretion disk (see, e.g., Refs.~\citep{Murase:2020lnu,Mummery2025}). Consequently, the detailed mechanism responsible for the thermal emission remains uncertain.
In addition, nonthermal emission components have been detected in some TDEs, particularly those associated with outflows. 
Radio emission in such cases is attributed to synchrotron radiation from relativistic electrons accelerated in shocks formed as jets or winds interact with the circumnuclear medium (CNM) (see, e.g., Ref.~\citep{Alexander2020}). 
Future multiwavelength observations are expected to provide crucial insights into the structure, energetics, and evolution of these outflows~\citep{DeColle2012,Alexander2020,van_Velzen2021}.

Some TDEs, such as Swift J1644+57~\citep{Bloom2011, Burrows2011}, Swift J2058+05~\citep{Cenko2012}, Swift J1112-8238~\citep{Brown2015}, and AT~2022cmc~\citep{Andreoni2022}, exhibit radio luminosities exceeding $10^{40}\,{\rm erg\;s^{-1}}$ and are widely interpreted as jetted TDEs, involving the launch of relativistic jets (e.g., Ref.~\citep{Giannios2011, Piran2015, van_Velzen2016, Matsumoto:2022hqp, Beniamini2023}).
In contrast, TDEs with lower peak radio luminosities of $\lesssim10^{38}$–$10^{39}\,{\rm erg\; s^{-1}}$ are generally attributed to sub-relativistic winds and are classified as jet-quiet TDEs (see, e.g., Refs.~\citep{Alexander2016, Matsumoto:2021qqo, Cendes2021, Hayasaki2023}).
Recently, however, some TDEs have been detected that exhibit radio flares peaking at late times (``delayed radio flare''), around $\sim10^8\;{\rm s}$ after their optical discovery, with peak luminosities of $\sim10^{38}$–$10^{39}\,{\rm erg\;s^{-1}}$ \citep{Horesh2021, Cendes2022, Cendes2023, Christy:2024szt}. 
Some of these delayed radio flares are preceded by another, earlier radio flare (``rebrightening radio events'').
These events are fainter than jetted TDEs, yet brighter than the typical jet-quiet TDE population at a few months. The origin of such delayed radio flares remains unclear.

Several models have been proposed to explain these delayed radio flares. 
One possibility is the delayed launch of an outflow, either as a spherical wind or a collimated relativistic jet, which interacts with the CNM at late times \citep{Horesh2021b, Cendes2023, Goodwin:2024cyg,Lin2025,Sfaradi2025,Golay2025,Wu2025}. 
Another possibility is outflow interaction with a stratified CNM, where the ambient density remains low initially but increases sharply at $\sim0.1-1$~pc, triggering strong radio emission once the outflow reaches this denser region~\citep{Dai2002, NG2007, Zhuang:2024tos}.
A third scenario involves a transition in the CNM density profile, from a power-law profile to a constant density medium beyond the Bondi radius~\citep{Matsumoto:2024eja}.
The fourth scenario is that TDE jets are launched promptly but remain choked by disk winds due to disk--spin misalignment, becoming observable after alignment with the SMBH spin axis \citep{Teboul:2023bpk}.
Finally, delayed radio brightening may also result from an off-axis relativistic jet. A one-component version of this scenario has been proposed by Ref.~\citep{Sfaradi2023}, while a two-component off-axis jet model has been suggested by Ref.~\citep{Sato2024}, in which a narrow fast jet and a wider, slower component contribute to the observed radio brightening. 
On the other hand, distinguishing among these possibilities remains challenging due to insufficient multiwavelength coverage during the late-time evolution of these TDEs \citep{Yuan2024,Yuan:2024sxk}.

In this paper, we investigate whether the radio emission observed in delayed radio flare TDEs can be consistently explained by delayed outflow models. 
We consider three scenarios: an instantaneous wind, a delayed wind, and a delayed relativistic jet.
To distinguish between these models, we examine the detectability of multiwavelength afterglow emission with the Cherenkov Telescope Array Observatory (CTAO) \citep{CTA}, the {\it Chandra} X-ray Observatory ({\it Chandra}) \citep{Chandra}, the Vera C. Rubin Observatory (Rubin) \citep{Rubin}, the Zwicky Transient Facility (ZTF) \citep{ZTF}, the Atacama Large Millimeter/submillimeter Array (ALMA) \citep{ALMA}, the Square Kilometre Array Observatory (SKAO) \citep{SKA}, the Very Large Array (VLA) \citep{VLA}, and the MeerKAT radio telescope (MeerKAT) \citep{MeerKAT}.
In particular, we assess whether x-ray and optical follow-up observations can help break the degeneracy between the models, which often produce similar radio light curves. By comparing model predictions with existing and forthcoming observations, we aim to constrain the physical origin of delayed radio flares and improve our understanding of outflow launch mechanisms in TDEs.
This paper is organized as follows.
In Section~\ref{sec:model}, we describe our modeling framework for the instantaneous wind, delayed wind, and delayed jet scenarios.
In Section~\ref{sec:result}, we compare the model predictions with observed radio light curves for a sample of TDEs exhibiting late-time radio emission.
In Section~\ref{sec:test}, we discuss the prospects of detecting associated multiwavelength (x-ray, optical, radio) afterglows with current and future facilities. 
Section~\ref{sec:discussion} is devoted to a discussion of the main results, and the conclusions are presented in Section~\ref{sec:conclusion}.

\section{Model description}
\label{sec:model}

\begin{figure*}
\vspace*{5pt}
\centering
\includegraphics[width=0.5\textwidth]{wind.pdf}
\caption{
Schematic illustration of the instantaneous wind model.  
A nonrelativistic wind is launched from the accretion disk after the stellar disruption. The wind drives a forward shock into the CNM with a power-law density profile, $n(R)$, producing synchrotron emission from shock accelerated electrons.
}
\label{fig:model_w}
\end{figure*}
\begin{figure*}
\vspace*{5pt}
\centering
\includegraphics[width=1.0\textwidth]{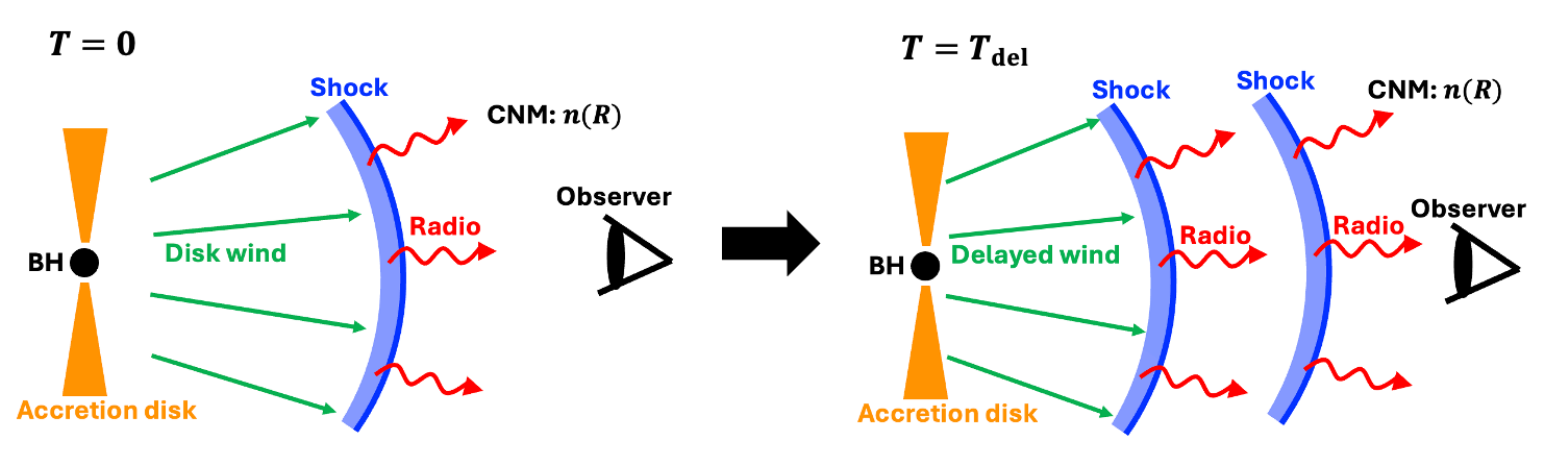}
\caption{
Schematic illustration of the delayed wind model.
At $T = 0$, the initial nonrelativistic wind is launched from the accretion disk. Subsequently, a second wind component is launched at a time delay $T = T_{\rm del}$, following the initial wind. Similar to the initial nonrelativistic wind, the delayed wind also interacts with the CNM, driving a shock that produces renewed synchrotron emission.
}
\label{fig:model_delayw}
\end{figure*}
\begin{figure*}
\vspace*{5pt}
\centering
\includegraphics[width=1.0\textwidth]{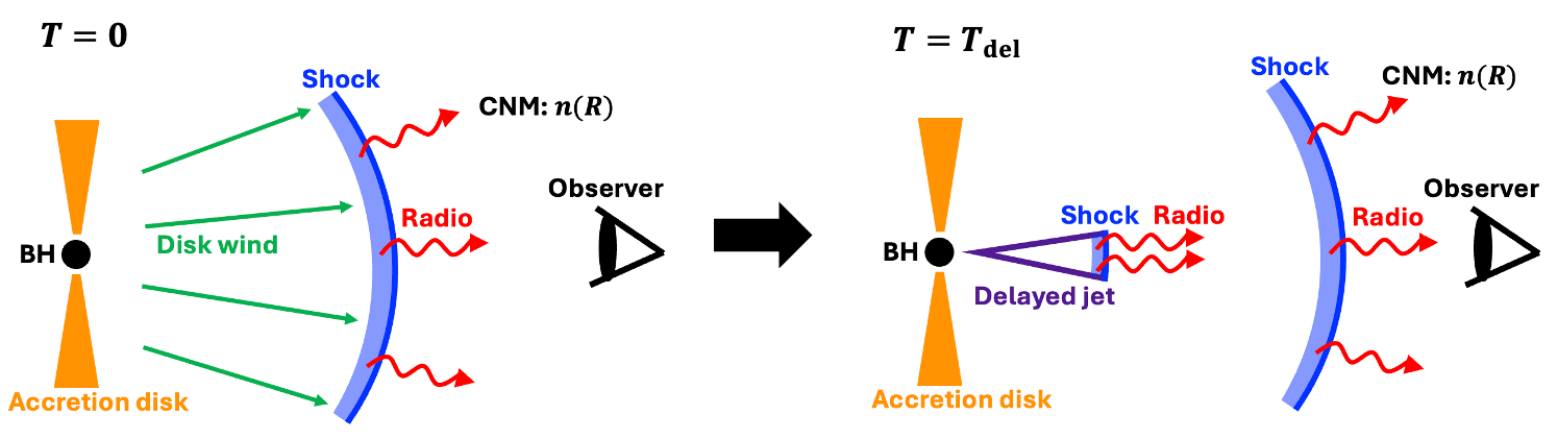}
\caption{
Schematic illustration of the delayed jet model. At $T = 0$, the initial nonrelativistic wind is launched from the accretion disk. Subsequently, at $T = T_{\rm del}$, a relativistic jet is launched. The jet propagates into the CNM, driving a forward shock that forms a thin shell. Synchrotron radiation is emitted from this shocked region as the jet decelerates and evolves.
}
\label{fig:model_delayj}
\end{figure*}

In this section, we describe three models, the instantaneous wind, delayed wind, and delayed jet, used to study the outflow dynamics and the associated synchrotron emission.
Our modeling framework follows the gamma-ray burst afterglow formalism, in which the outflow dynamics are evolved and synchrotron and inverse-Compton emission is computed.
To compute the resulting radio light curves, we utilize the blast wave module of {\sc AMES} (Astrophysical Multimessenger Emission Simulator), following the methodology described in Refs.~\citep{Zhang2021,Zhang:2023uei} (see also Refs.~\citep{Murase:2010fq,Murase:2023chr,Sato2024}).
The blast-wave evolution and the associated radiation are calculated numerically in {\sc AMES}, which follows the dynamical evolution of the outflow from the coasting phase through deceleration, including both relativistic and nonrelativistic outflows and, at sufficiently late times, the deep-Newtonian phase.
The analytic expressions presented below are not used in the numerical calculations, but are introduced to facilitate the interpretation of the numerical results and to illustrate the asymptotic parameter scalings.

\subsection{Instantaneous wind model}
\label{sec:wind}
We model the dynamical evolution of a spherical nonrelativistic outflow characterized by an isotropic-equivalent kinetic energy ${\mathcal E}_{k}^{\rm iso}$ and an initial velocity $v_0$.
The outflow propagates into the CNM described by a power-law density profile, $n(R) = AR^{-w} = n_{\rm ext} (R/10^{18}\,{\rm cm})^{-w}$~(see Fig.~\ref{fig:model_w} for a schematic picture).
We assume that the emission originates from a region at radius $R$.
The outflow enters the deceleration phase, for which the radius and velocity approach the self-similar evolution given by
\footnotesize
\begin{eqnarray}
\label{eq:non-rela}
    R &=& \left(1+z\right)^{-\frac{2}{5-w}}\left(\frac{5-w}{2}\right)^{\frac{2}{5-w}}\left(\frac{(3-w){\mathcal E}_{k}^{\rm iso}}{2\pi Am_p}\right)^{\frac{1}{5-w}}T^{\frac{2}{5-w}}, \\
    v &=& \left(1+z\right)^{-\frac{w-3}{5-w}}\left(\frac{5-w}{2}\right)^{\frac{w-3}{5-w}}\left(\frac{(3-w){\mathcal E}_{k}^{\rm iso}}{2\pi A m_p}\right)^{\frac{1}{5-w}}T^{\frac{w-3}{5-w}},
\end{eqnarray}
\normalsize
where $v$ is the outflow velocity, $m_p$ is the proton mass, and $T = t - t_0$ is the observer time measured from the time of optical discovery ($t_0$), at which the tidal disruption occurs and the outflow is assumed to be launched (see, e.g., Refs.~\citep{Wijers:1997xu,Dai:1999fc,Huang:2003nw}).
As the outflow interacts with the dense external medium, synchrotron and synchrotron self-Compton (SSC) emission is generated.
We adopt a power-law energy distribution for the accelerated electrons with index $s$, and assume constant microphysical parameters $\epsilon_B$, $\epsilon_e$, and $f_e$, which represent the fractions of internal energy transferred to magnetic fields, nonthermal electrons, and the fraction of accelerated electrons, respectively.
The downstream magnetic field, amplified by shocks, is estimated as (e.g., Refs.~\cite{Piran:2004ba,Bing2018})
\begin{equation}
    B' = \left(8\pi \epsilon_B  m_p n(R) v^2\right)^{1/2}.
\end{equation}
The minimum Lorentz factor of accelerated electrons is
\begin{equation}
    \gamma_m \simeq \frac{\epsilon_e}{f_e} \left(\frac{s-2}{s-1}\right) \frac{m_p}{m_e} \frac{v^2}{c^2}.
\end{equation}
The corresponding minimum synchrotron frequency is 
\begin{align}
\label{nu_m}
    \nu_m \approx& (5.1\times10^{8}\, {\rm Hz})\, \left(1+z\right)^{\frac{10-3w}{5-w}}\left(\frac{s-2}{s-1}\right)^2 \mathcal{E}_{52}^{\frac{10-w}{2(5-w)}}  \nonumber\\
    &\times  n_{-1}^{-\frac{5}{2(5-w)}}\epsilon_{B,-2}^{\frac{1}{2}} \epsilon_{e,-1}^{2} f_{e,0}^{-2} T_7^{\frac{4w-15}{5-w}},
\end{align}
where $z$ is the redshift, $\mathcal{E}_{52}={\mathcal E}_{k}^{\rm iso}/10^{52}\,{\rm erg}$, $n_{-1}=n_{\rm ext}/10^{-1}\,{\rm cm^{-3}}$, $\epsilon_{B,-2}=\epsilon_{B}/10^{-2}$, $\epsilon_{e,-1}=\epsilon_{e}/10^{-1}$, $f_{e,0}=f_{e}/1.0$, and $T_7=T/10^7\,{\rm s}$.
The synchrotron cooling frequency is given by
\footnotesize
\begin{equation}
\label{nu_c}
    \nu_c \approx (5.6\times10^{15}\, {\rm Hz})\, \left(1+z\right)^{\frac{7-w}{5-w}} \mathcal{E}_{52}^{-\frac{3(2-w)}{2(5-w)}} n_{-1}^{-\frac{3(7-2w)}{2(5-w)}} \epsilon_{B,-2}^{-\frac{3}{2}} T_7^{\frac{2w-1}{5-w}}.
\end{equation}
\normalsize
Synchrotron self-absorption suppresses radio emission at low frequencies.
The synchrotron self-absorption frequency $\nu_a$ is defined by the condition $\tau_{\rm ssa}(\nu_a) = 1$, where $\tau_{\rm ssa}$ is the corresponding optical depth.

As the observed radio spectra of most TDEs exhibit the frequency ordering $\nu_m < \nu_a < \nu_c$ at late times (see, e.g., Ref.~\citep{Cendes2023}), we adopt the same spectral regime to compute the synchrotron self-absorption frequency:
\begin{align}
\label{nu_a}
    \nu_a \approx& (1.1\times10^{9}\, {\rm Hz})\, \left(1+z\right)^{\frac{3(4-s)w+10(s-3)}{(5-w)(4+s)}} \left(\frac{s-2}{s-1}\right)^{\frac{2(s-1)}{s+4}}  \nonumber\\ 
    &\times \mathcal{E}_{52}^{\frac{10s-(s+6)w}{2(5-w)(s+4)}} n_{-1}^{\frac{5(6-s)}{2(5-w)(s+4)}} \epsilon_{B,-2}^{\frac{s+2}{2(s+4)}} \epsilon_{e,-1}^{\frac{2(s-1)}{s+4}}  f_{e,0}^{\frac{2(2-s)}{s+4}}\nonumber\\
    &\times T_7^{\frac{4(s-2)w+5(2-3s)}{(5-w)(s+4)}}.
\end{align}
The synchrotron peak flux is given by
\begin{align}
\label{Fnu_max}
    F_{\nu}^{\rm max} \approx& (1.7\, {\rm mJy})\, \left(1+z\right)^{\frac{w+2}{5-w}} d_{L,28}^{-2} \mathcal{E}_{52}^{\frac{8-3w}{2(5-w)}} n_{-1}^{\frac{7}{2(5-w)}} \nonumber\\
    &\times\epsilon_{B,-2}^{\frac{1}{2}} f_{e,0} T_7^{\frac{3-2w}{5-w}},
\end{align}
where $d_L = d_{L,28} \times 10^{28}\,{\rm cm}$ is the luminosity distance.
The synchrotron spectrum in the regime $\nu_m < \nu_a < \nu_c$ takes the form
\begin{equation}
\label{Fnu_total}
    F_{\nu} = F_{\nu}^{\rm max} \begin{cases}
    \left(\frac{\nu}{\nu_m}\right)^2 \left(\frac{\nu_a}{\nu_m}\right)^{-\frac{s}{2}-2}, & \nu \leq \nu_m,\\
    \left(\frac{\nu}{\nu_a}\right)^{\frac{5}{2}} \left(\frac{\nu_a}{\nu_m}\right)^{-\frac{s-1}{2}}, & \nu_m < \nu \leq \nu_a,\\
    \left(\frac{\nu}{\nu_m}\right)^{-\frac{s-1}{2}}, & \nu_a < \nu \leq \nu_c,\\
    \left(\frac{\nu}{\nu_c}\right)^{-\frac{s}{2}} \left(\frac{\nu_c}{\nu_m}\right)^{-\frac{s-1}{2}}, & \nu_c < \nu.
    \end{cases}
\end{equation}
Under the condition $\nu_m \leq \nu \leq \nu_a \leq \nu_c$, substituting $\nu_m$, $\nu_a$, and $F_{\nu}^{\rm max}$ into Eq.~\ref{Fnu_total} yields the observed flux as a function of time:
\begin{align}
\label{Fnu_degen}
    F_\nu(T) \sim& (3.8\times10^{-1}\, {\rm mJy})\, \left(1+z\right)^{\frac{24-7w}{2(5-w)}}\mathcal{E}_{52}^{\frac{w+6}{4(5-w)}} n_{-1}^{-\frac{11}{4(5-w)}} \nonumber\\
    &\times \epsilon_{B,-2}^{-\frac{1}{4}} d_{L,28}^{-2}\nu_9^{\frac{5}{2}} T_7^{\frac{11}{2(5-w)}},
\end{align}
where $\nu_9 = \nu / 10^9\,{\rm Hz}$.
These scalings of Eqs.~(\ref{nu_m}), (\ref{nu_a}), and (\ref{Fnu_degen}) are consistent with Ref.~\citep{Leventis2012}.
In this regime, the light curve peaks when the synchrotron self-absorption frequency $\nu_a$ crosses the observing frequency.
After the peak, the flux decays as $F_\nu \propto T^{\frac{(s-2)4w+3(7-5s)}{2(5-w)}}$.

In the above analytic treatment, the accelerated electrons are assumed to remain relativistic, with $\gamma_m\propto (v/c)^2$. 
At sufficiently late times, the blast wave can enter the deep-Newtonian phase, in which the minimum Lorentz factor of accelerated electrons contributing to the synchrotron emission is approximately constant (e.g. Ref.~\citep{Sironi:2013tva}). 
In this phase, the downstream magnetic field and the minimum synchrotron frequency scale as $B'\propto T^{-3/(5-w)}$ and $\nu_m\propto T^{-3/(5-w)}$, respectively, while the peak flux scales as $F_\nu^{\rm max}\propto T^{-3/(5-w)}$. 
For $\nu_m<\nu_a<\nu<\nu_c$, the expected late-time decay in the deep-Newtonian phase becomes $F_\nu\propto T^{-\frac{3(s+1)}{2(5-w)}}$ (e.g. Ref.~\citep{Sironi:2013tva}).

Before the onset of deceleration, the outflow undergoes a coasting phase with $R\propto T$ and $v\simeq v_0={\rm const.}$. The deceleration time, defined by the condition that the swept-up circum-nuclear material becomes comparable to the ejecta mass, is approximately
$T_{\rm dec}\simeq(1+z)\left[\frac{(3-w){\mathcal E}_{k}^{\rm iso}}{2\pi A m_p}\right]^{1/(3-w)}v_0^{-\frac{5-w}{3-w}}$ (e.g. Ref.~\citep{Minhajur2026}).
For $T\lesssim T_{\rm dec}$, the relevant synchrotron quantities scale as
$B’\propto T^{-w/2}$,
$\nu_m\propto T^{-w/2}$,
$\nu_a\propto T^{\frac{4-w(6+s)}{2(s+4)}}$, and
$F_{\nu}^{\rm max}\propto T^{\frac{3(2-w)}{2}}$.
The corresponding synchrotron flux evolves as
$F_\nu\propto T^{\frac{8+w}{4}}$
for $\nu_m<\nu<\nu_a<\nu_c$
and
$F_\nu\propto T^{\frac{12-w(s+5)}{4}}$
for $\nu_m<\nu_a<\nu<\nu_c$.

\subsection{Delayed wind model}
\label{sec:delayed-wind}
In this model, the second delayed outflow is launched from the accretion disk at a delay time $T_{\rm del}$ following the initial nonrelativistic outflow (see Fig.~\ref{fig:model_delayw}).
This delayed wind, characterized by an isotropic kinetic energy ${\mathcal E}_{k}^{\rm iso}$ and an initial velocity $v_0$, interacts with the CNM, giving rise to renewed synchrotron emission from shock accelerated electrons. The electron energy distribution is assumed to follow a power-law with index $s$, and the emission is characterized by microphysical parameters $\epsilon_e$, $\epsilon_B$, and $f_e$. The outflow dynamics and synchrotron radiation are modeled in the same manner as in Section~\ref{sec:wind}, but with a shifted time origin.

The delayed onset compresses the emission duration in the observer time frame, resulting in a more rapid flux rise shortly after the launch of the delayed outflow.
Furthermore, the temporal evolution of the synchrotron break frequencies ($\nu_a$, $\nu_m$, and $\nu_c$) and the flux normalization is governed by the shifted time coordinate $T - T_{\rm del}$.
As a result, simple analytical power-law scalings in the observer time $T$ are generally not applicable.
Nevertheless, similar to the instantaneous wind case, the light curve is still expected to peak when the synchrotron self-absorption frequency $\nu_a$ passes through the observing band.

\subsection{Delayed jet model}
\label{sec:delayed-jet}

In the delayed jet model, we assume that a relativistic jet is launched from the SMBH at a delay time $T_{\rm del}$, following the initial nonrelativistic wind as shown in Fig.~\ref{fig:model_delayj}. 
The jet propagates into the CNM with a density profile $n(R) = n_{\rm ext}(R/10^{18}\;{\rm cm})^{-w}$ and is modeled as a top-hat jet with isotropic-equivalent kinetic energy $\mathcal{E}_{k}^{\rm iso}$, initial bulk Lorentz factor $\Gamma_0$, and initial half-opening angle $\theta_0$.
As the jet expands, it drives a forward shock into the CNM, forming a thin shell that produces synchrotron emission. The radiative treatment follows the same prescription as discussed in Section~\ref{sec:wind}. We assume an on-axis observer for our analysis. 

As with the delayed wind model, the time shift due to $T_{\rm del}$ compresses the light curve in the observer time frame, making simple power-law approximations with $T$ generally invalid.
On the other hand, the light curve is still expected to peak when the synchrotron self-absorption frequency $\nu_a$ crosses the observing frequency.

\begin{table*}
\caption{Model fit summary for TDEs with detected radio emission}
\label{tab:fit}
\renewcommand{\arraystretch}{1.2}
\scalebox{1.2}{
\begin{tabular}{lcccc}
\hline
\textbf{TDE Name}~~ & ~~\textbf{Two outflows}${}^{\rm a}$~~ & ~~\textbf{Instantaneous wind}~~ & ~~\textbf{Delayed wind}${}^{\rm b}$~~ & ~~\textbf{Delayed jet}${}^{\rm c}$~~ \\
\hline
ASASSN-14ae${}^{\rm d}$ & Yes & $\times$ & \checkmark & \checkmark \\
ASASSN-15oi${}^{\rm d,f}$ & Yes &  $\times$ & \checkmark & \checkmark \\
PS16dtm${}^{\rm d}$ & Yes & $\times$ & \checkmark & \checkmark \\
iPTF16fnl${}^{\rm e}$ &  &  \checkmark & \checkmark & $\times$ \\
ASASSN-19bt${}^{\rm d,f}$ & Yes & $\times$ & \checkmark & $\times$ \\
AT~2019dsg${}^{\rm d,f}$ & Yes & $\times$ & \checkmark & \checkmark \\
AT~2019ehz${}^{\rm d}$ &  & \checkmark & \checkmark & \checkmark \\
AT~2019eve${}^{\rm d}$ &  & \checkmark & \checkmark & \checkmark \\
AT~2020vwl${}^{\rm d,f}$ & Yes & $\times$ & \checkmark & $\times$ \\
\hline
\end{tabular}}
\renewcommand{\arraystretch}{1}
\begin{tablenotes}
\item {\bf Notes.} \\
${}^{\rm a}$~Two separate outflows (instantaneous wind+delayed wind and/or instantaneous wind+delayed jet).\\
${}^{\rm b}$~Instantaneous wind+delayed wind.\\
${}^{\rm c}$~Instantaneous wind+delayed jet.\\
${}^{\rm d}$~Delayed radio flare event.\\
${}^{\rm e}$~Radio event without a delayed flare.\\
${}^{\rm f}$~Rebrightening radio event.\\
\end{tablenotes}
\end{table*}

\section{Comparison of theoretical models with observations}
\label{sec:result}

\begin{figure*}
\vspace*{5pt}
\begin{minipage}{0.45\linewidth}
\centering
\includegraphics[width=1.0\textwidth]{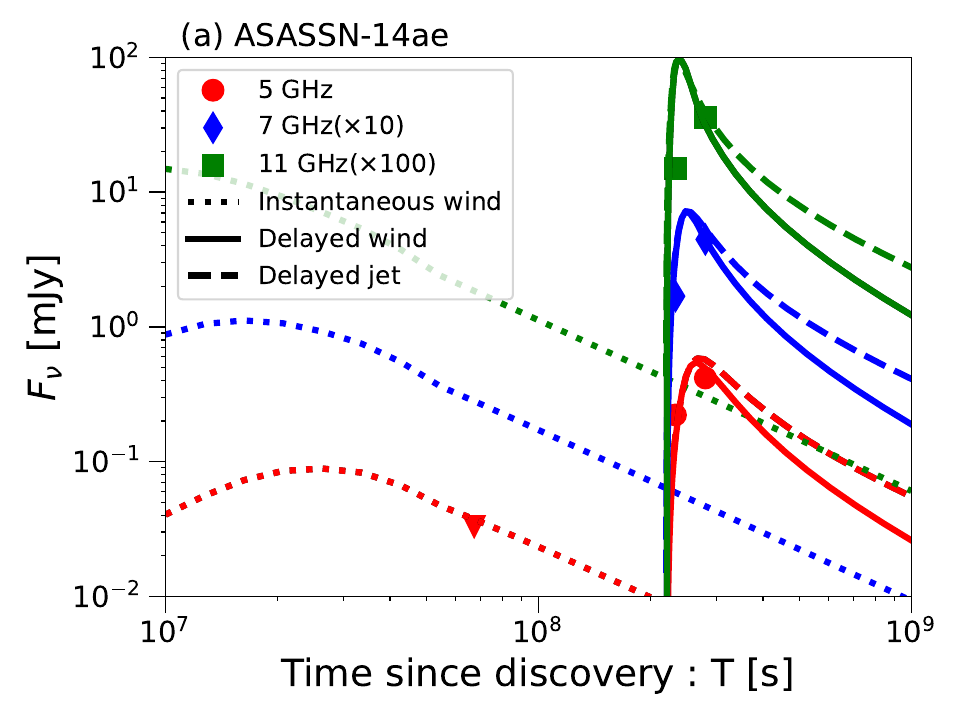}
\end{minipage}
\begin{minipage}{0.45\linewidth}
\centering
\includegraphics[width=1.0\textwidth]{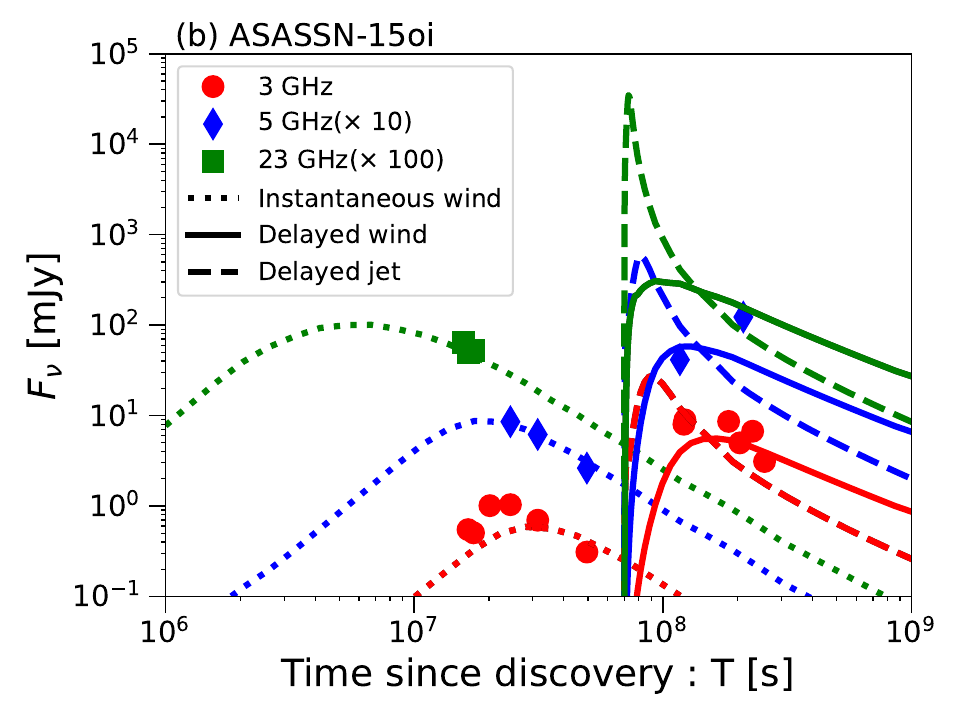}
\end{minipage}
\begin{minipage}{0.45\linewidth}
\centering
\includegraphics[width=1.0\textwidth]{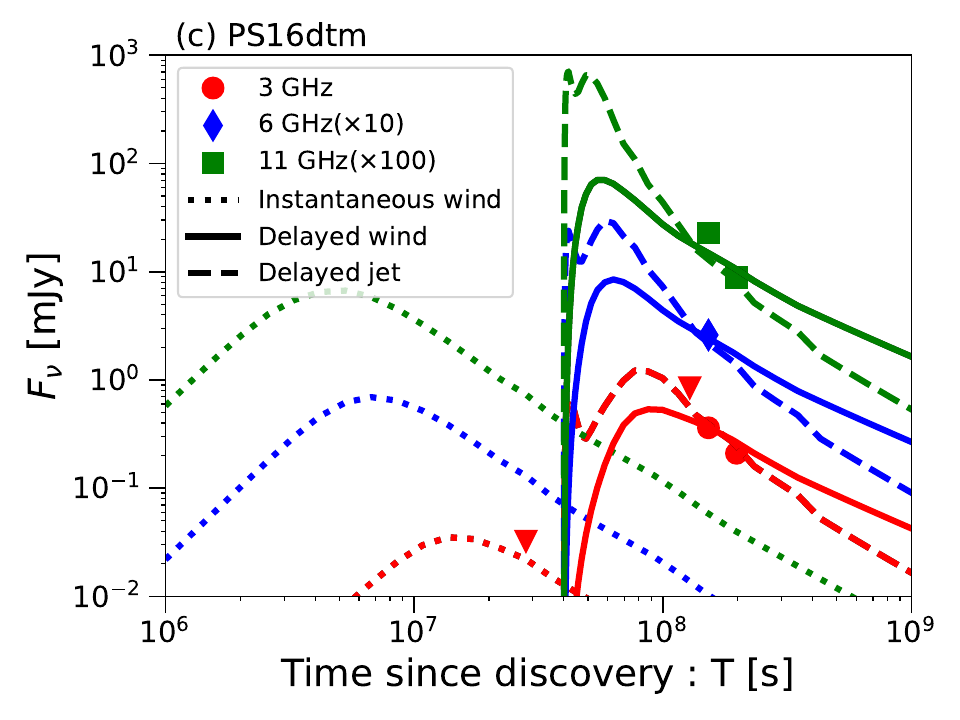}
\end{minipage}
\begin{minipage}{0.45\linewidth}
\centering
\includegraphics[width=1.0\textwidth]{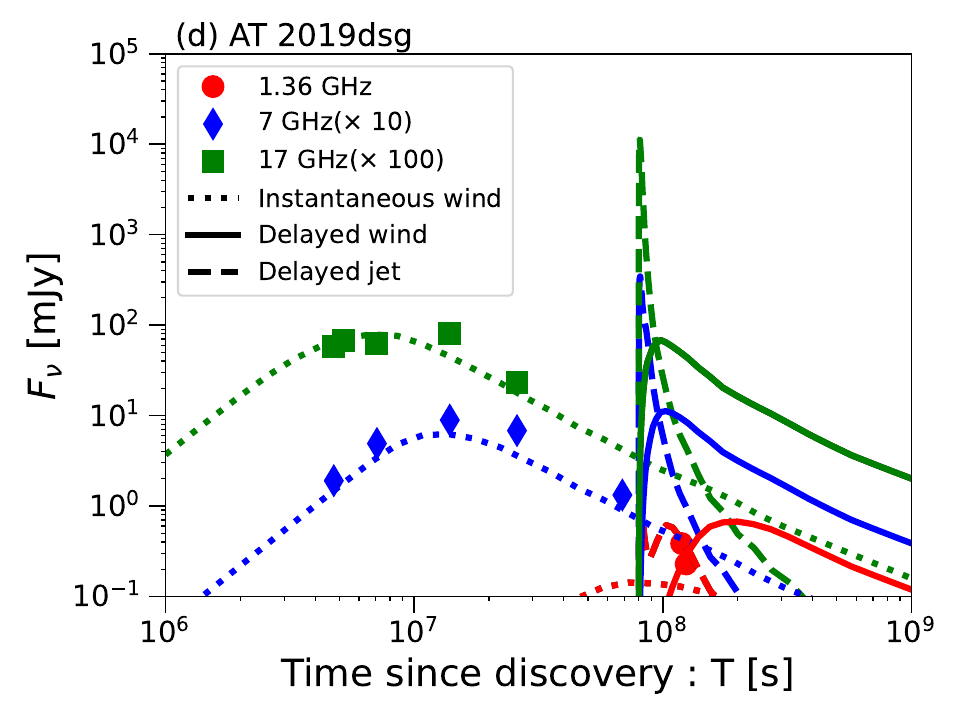}
\end{minipage}
\vspace{-0.3cm}
\caption{
Observed radio data for ASASSN-14ae in panel (a), ASASSN-15oi in panel (b), PS16dtm in panel (c), and AT~2019dsg in panel (d) are shown (see plot legends for frequencies corresponding to radio observations considered). These are compared with the radio light curves calculated from our theoretical models. In each panel, the dotted, solid, and dashed lines represent the emission from the instantaneous wind, delayed wind and delayed jet, respectively. The upper limits in radio flux are shown with downward triangles. For these four events, the observed radio data suggests the presence of delayed outflows, either in the form of a jet or a wind.
}
\label{fig:result1}
\end{figure*}
\begin{figure*}
\vspace*{5pt}
\begin{minipage}{0.45\linewidth}
\centering
\includegraphics[width=1.0\textwidth]{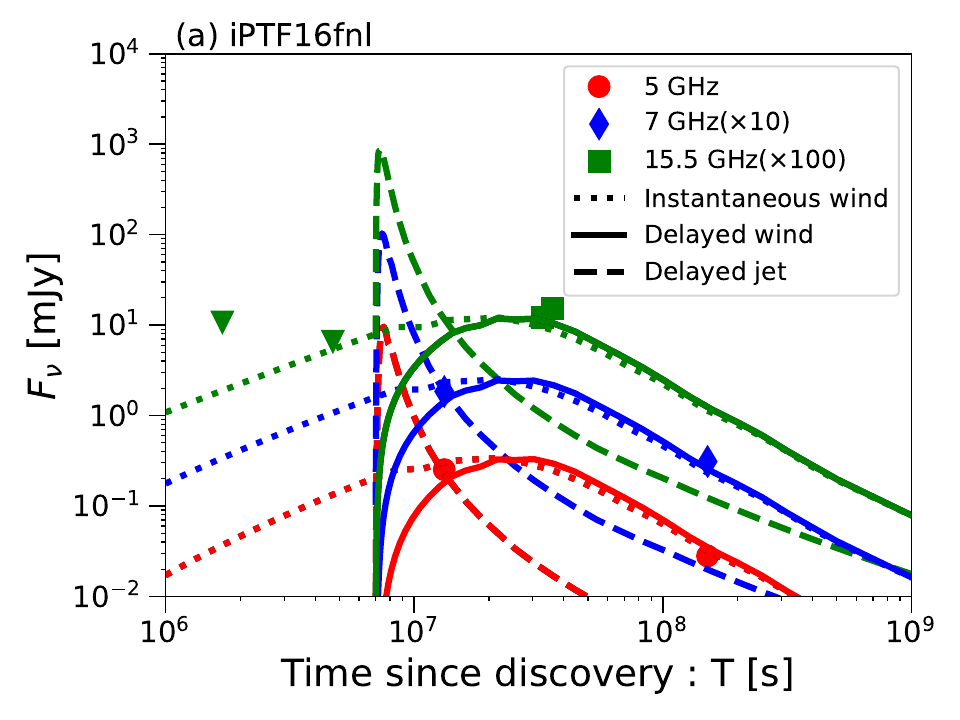}
\end{minipage}
\begin{minipage}{0.45\linewidth}
\centering
\includegraphics[width=1.0\textwidth]{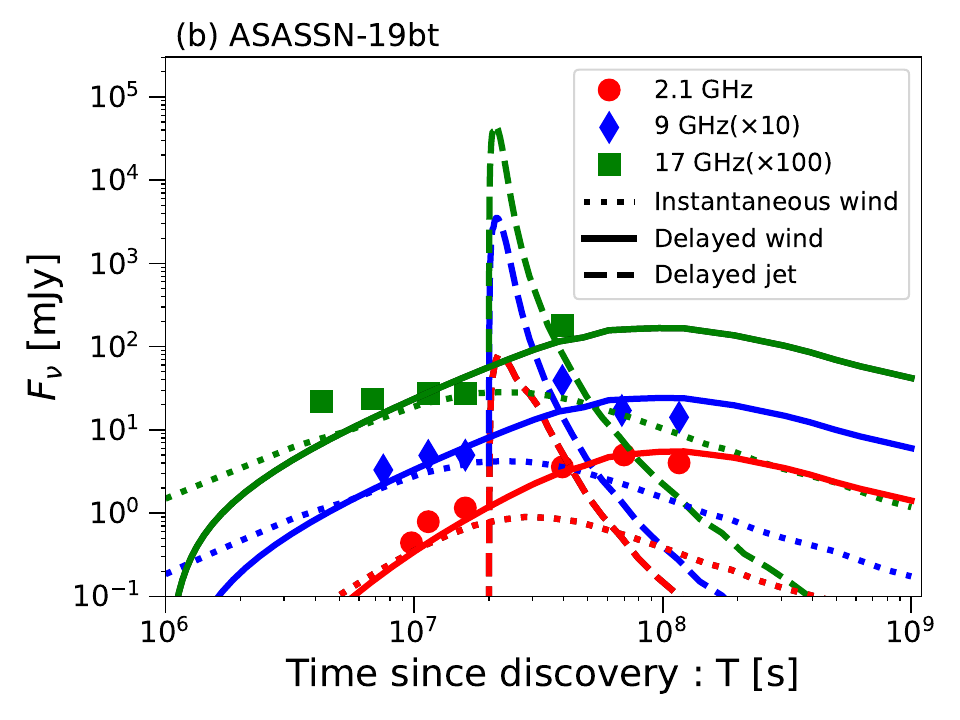}
\end{minipage}
\begin{minipage}{0.45\linewidth}
\centering
\includegraphics[width=1.0\textwidth]{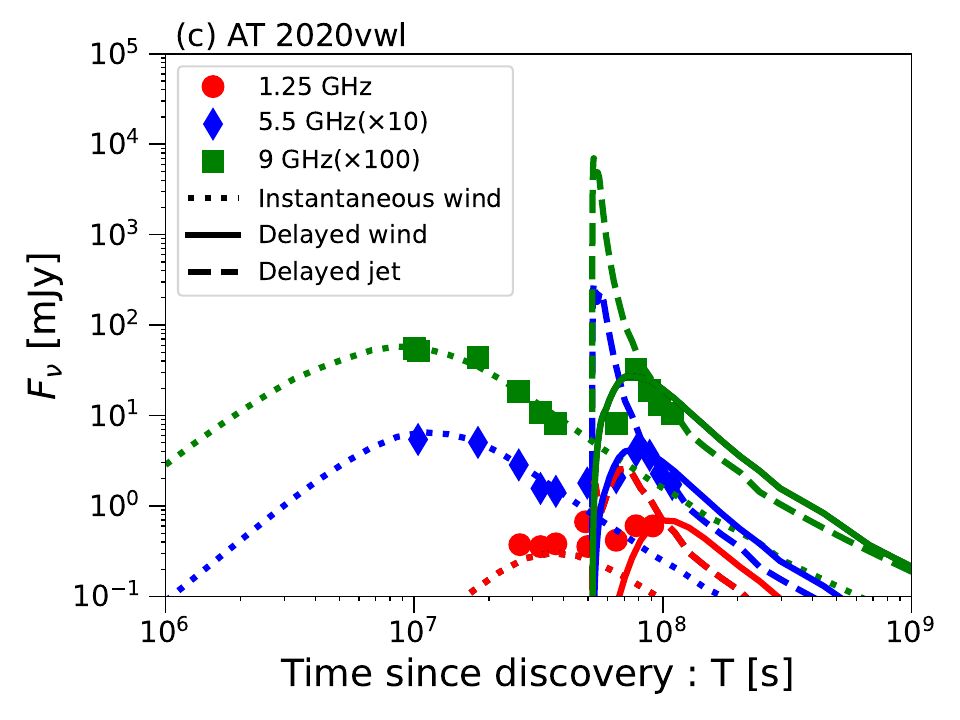}
\end{minipage}
\vspace{-0.3cm}
\caption{
Observed radio data for iPTF16fnl in panel (a), ASASSN-19bt in panel (b), and AT~2020vwl in panel (c) are shown (see plot legends for frequencies corresponding to radio observations considered). The radio observations are compared with the predictions of our theoretical models. The line types and downward triangles shown here have the same meaning as in Figure~\ref{fig:result1}.
For these three events, the radio emission from a delayed jet is inconsistent with the observed radio flux.
}
\label{fig:result2}
\end{figure*}
\begin{figure*}
\vspace*{5pt}
\begin{minipage}{0.45\linewidth}
\centering
\includegraphics[width=1.0\textwidth]{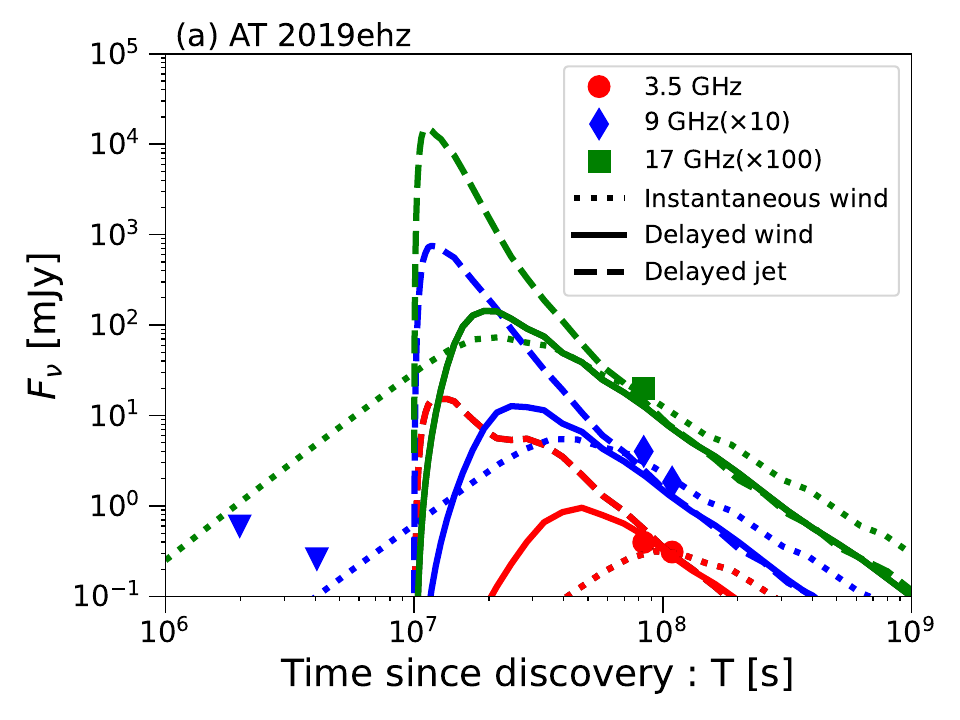}
\end{minipage}
\begin{minipage}{0.45\linewidth}
\centering
\includegraphics[width=1.0\textwidth]{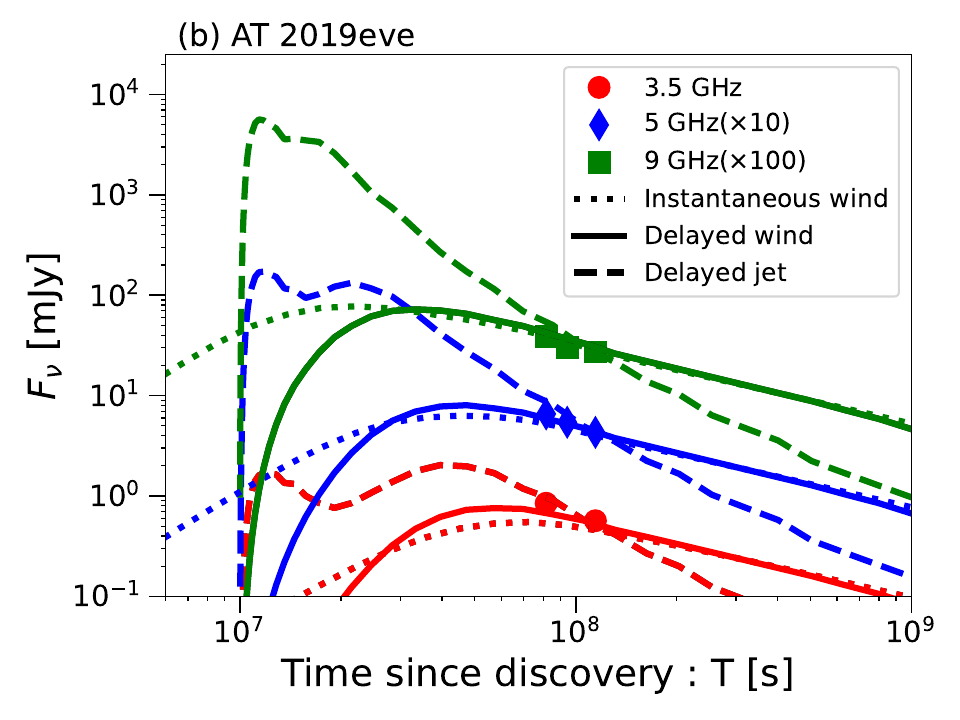}
\end{minipage}
\vspace{-0.3cm}
\caption{
Observed radio data for AT~2019ehz in panel (a) and AT~2019eve in panel (b) are shown (see plot legends for corresponding observing frequencies). Radio observations are compared with light curves inferred from the three theoretical models. The line types and downward triangles have the same meaning as in Figure~\ref{fig:result1}.
For these events, all three models, the instantaneous wind, delayed wind, and delayed jet, can explain the observed radio data.
}
\label{fig:result3}
\end{figure*}

We model the radio light curves of TDEs using three scenarios: (i) an instantaneous wind model, (ii) a delayed wind model, and (iii) a delayed jet model.
Our sample consists of 9 TDEs with well sampled radio light curves, namely ASASSN-14ae, ASASSN-15oi, PS16dtm, iPTF16fnl, ASASSN-19bt, AT~2019dsg, AT~2019ehz, AT~2019eve, and AT~2020vwl.
Several events in our sample overlap with those analyzed by Ref.~\citep{Wu2025}.

We classify an event as a delayed radio flare if its peak radio luminosity exceeds $5 \times 10^{37}~{\rm erg\,s^{-1}}$ at $T > 5 \times 10^{6}~{\rm s}$. 
Based on this criterion, eight events (all apart from iPTF16fnl) are classified as delayed radio flares. A summary of references reporting radio observations for these sources is provided in Table~\ref{tab:reference}.
To compute the radio light curves, we utilize the {\sc AMES} code (see Section~\ref{sec:model}). {\sc AMES} incorporates key physical processes, including outflow dynamics, synchrotron radiation mechanisms and absorption processes.

From our analysis, we find that the majority of sources are well described by the following typical parameter ranges: 
$n_{\rm ext} \sim 10$--$10^3~{\rm cm^{-3}}$, 
$w \sim 1$--$2$,
$s \sim 2.1$--$2.8$, 
$\epsilon_B \sim 10^{-3}$--$10^{-2}$, 
$\epsilon_e \sim 10^{-3}$--$0.1$, and 
$f_e \sim 0.1$. 
 A few sources exhibit parameters outside these ranges, such as ASASSN-14ae ($n_{\rm ext}\sim10^{5}\,{\rm cm^{-3}}$), iPTF16fnl ($\epsilon_B\sim3\times10^{-5}$), and AT~2019dsg ($\epsilon_B\sim10^{-4}$). These deviations likely reflect parameter degeneracies inherent to the model.
For the wind component, we adopt $v_0 \sim 0.1c$ and ${\mathcal E}_{k}^{\rm iso} \sim 5 \times 10^{51}-5 \times 10^{52}~{\rm erg}$. For the relativistic jet component, we assume $\Gamma_0 \sim 3$, $\theta_0 \sim 0.35~{\rm rad}$, and ${\mathcal E}_{k}^{\rm iso} \sim 5 \times 10^{52}~{\rm erg}$. 
The launch time of the delayed components is set to $T_{\rm del} \sim 10^7$--$10^8~{\rm s}$.
The best fit radio light curves are shown in Figures~\ref{fig:result1}, \ref{fig:result2}, and \ref{fig:result3}, and Table~\ref{tab:fit}, with the best fit model parameters summarized in Table~\ref{tab:parameter}.

For the instantaneous wind model, most events remain in the coasting phase over the epochs relevant to the observed radio emission, with typical deceleration times of $T_{\rm dec}\sim10^8$–$10^9\,{\rm s}$.
Only ASASSN-14ae has entered the deceleration phase.
Accordingly, the temporal evolution of the synchrotron emission for most sources is governed by the coasting-phase dynamics rather than by the asymptotic self-similar scalings of the deceleration phase.
The numerical light curves shown in this work account for the dynamical evolution directly through the time-dependent calculation implemented in {\sc AMES}.

The physical parameters adopted in this work are broadly informed by those inferred from radio observations of TDEs in previous studies, in particular Ref.~\citep{Cendes2023}, while reflecting differences in modeling assumptions.
Ref.~\citep{Cendes2023} employs an equipartition analysis of late-time radio spectral energy distributions, which yields characteristic kinetic energies of ${\sim}10^{47}$--$10^{49}$ erg, mildly sub-relativistic velocities ($v_0 \sim 10^{-2}$--$10^{-1}~c$), and ambient densities spanning a wide range across different events ($n_{\rm ext} \sim 1$--$10^{3}\,{\rm cm^{-3}}$).
In contrast, in our model, we explicitly solve time-dependent hydrodynamical evolution of winds and jets, and adopt microphysical parameters that are similar to those of gamma-ray burst afterglows. The inferred values are physically motivated and differ from those obtained by the equipartition assumption, as they correspond to different definitions and physical assumptions, including geometry and radiative efficiency. 
For example, we find isotropic-equivalent kinetic energies of ${\mathcal E}_{k}^{\rm iso} \sim 10^{51}$--$10^{52}$ erg. 
Although the kinetic energies inferred in our model are several orders of magnitude larger than those obtained in Ref.~\citep{Cendes2023}, the inferred CNM density scale remains broadly similar.
We note that Ref.~\citep{Cendes2023} primarily considers nonrelativistic wind scenarios, with delayed components discussed in detail for AT~2019dsg, whereas our analysis systematically examines both instantaneous and delayed winds as well as delayed relativistic jets across the TDE sample.

\begin{table}
\caption{References for the observed data are listed.}
\label{tab:reference}
\renewcommand{\arraystretch}{1.3}
\scalebox{1.0}{
\begin{tabular}{lll}
\hline
Events & Frequencies & references \\
\hline
ASASSN-14ae~& 5,~7,~and~11 GHz~ & Ref.~\citep{Cendes2023} \\
ASASSN-15oi~& 3,~5,~and~23~GHz~ &  Refs.~\citep{Horesh2021,Anumarlapudi2024,Hajela2024} \\
PS16dtm~ & 3,~6,~and~11~GHz~ & Ref.~\citep{Cendes2023} \\
iPTF16fnl~ & 5,~7,~and~15.5~GHz~ & Refs.~\citep{Horesh:2021szq,Cendes2023} \\
ASASSN-19bt~ & 2.1,~9,~and~17~GHz~ & Ref.~\citep{Christy:2024szt} \\
AT~2019dsg~ & 1.36,~7,~and~17~GHz~ & Refs.~\citep{Cendes2021,Cendes2023} \\
AT~2019ehz~& 3.5,~9,~and~20~GHz~ &  Ref.~\citep{Cendes2023} \\
AT~2019eve~ & 3.5,~5,~and~9~GHz~ & Ref.~\citep{Cendes2023} \\
AT~2020vwl~ & 1.25,~5.5,~and~9~GHz~ & Refs.~\citep{Goodwin:2023qcq,Goodwin:2024cyg} \\
\hline
\end{tabular}}
\renewcommand{\arraystretch}{1.0}
\end{table}

\subsection{TDEs with delayed radio flares}
\label{sub:delayed_TDE}

For the following events, the instantaneous wind scenario     struggles to explain the late-time rise in the radio flux, as it predicts a monotonic decline once the absorption frequency $\nu_a$ passes through the radio bands (see Sec.~\ref{sec:wind}). 
Instead, the delayed wind and/or delayed jet models, in which the outflow is launched at $T_{\rm del} \sim 10^7$–$10^8\,{\rm s}$, can account for the delayed radio brightening observed in these events:

{\it ASASSN-14ae, ASASSN-15oi, PS16dtm, \& AT~2019dsg.}  
In the instantaneous wind model, the radio flux decays after $\nu_a$ crosses the observational band, which is inconsistent with the observed late-time rise. A delayed wind or jet scenario, where the outflow is launched after a significant time lag ($T_{\rm del} \sim 10^7$–$10^8\,{\rm s}$), can explain the delayed peak in the radio light curve.
For PS16dtm, the light curve in the delayed jet scenario exhibits two distinct peaks. The first peak corresponds to the deceleration time of the jet, while the second arises when the synchrotron self-absorption frequency crosses the observing frequency.

{\it ASASSN-19bt \& AT~2020vwl.}  
Similar to ASASSN-15oi and AT~2019dsg, these two events exhibit detectable radio emission not only at early times but also at late times ($\sim10^8$~s).
In particular, AT~2020vwl displays rebrightening in its radio light curve. Delayed radio rebrightening is difficult to reconcile with a single outflow component and instead suggests the presence of an additional delayed outflow.
The late-time radio emission in these events is consistent with the delayed wind model, whereas the delayed jet scenario is disfavored. This is because, in the Blandford–McKee phase, the relativistic shock decelerates rapidly ($R \propto T^{1/(4-w)}$, $\Gamma \propto T^{(w-3)/2(4-w)}$), leading to a sharp post peak decline in emission. In contrast, a Newtonian outflow decelerates more gradually ($v \propto T^{(w-3)/(5-w)}$), producing a slower flux decay, consistent with the observed light curves.

{\it AT~2019ehz \& AT~2019eve.}  
The observed radio light curve can be explained by either an instantaneous wind or a delayed outflow (wind or jet). In these cases, the introduction of a delayed component is not strictly necessary.
For AT~2019eve, the light curve in the delayed jet scenario exhibits two distinct peaks. While the first peak corresponds to the deceleration time of the jet, the second emerges when the synchrotron self-absorption frequency drops below the observing frequency, similar to the behavior observed in PS16dtm.

\subsection{TDEs without delayed radio flares}
\label{sub:non-delayed_TDE}

{\it iPTF16fnl.}  
Both the instantaneous wind and delayed wind models can reproduce the late-time radio data.
Note that the inferred isotropic kinetic energy using the delayed wind model is lower than that of the delayed flare TDEs. This is because iPTF16fnl is the only event in our sample classified as exhibiting a TDE without delayed radio flare, whose peak radio luminosity is lower than that of the events with delayed flares.
In contrast, the delayed jet model predicts a rapid decline after the peak, which is inconsistent with the observed radio light curve. As in ASASSN-19bt and AT~2020vwl, this arises due to deceleration of the relativistic jet, which causes a steep decline in radio emission as the relativistic beaming reduces. In contrast, emission from the nonrelativistic wind decays more gradually, as the outflow remains in the Newtonian regime. Given that the instantaneous wind model requires fewer model parameters compared to the delayed wind or jet scenarios, it is favored as a natural explanation for this event.

\section{Testing models with multiwavelength observations}
\label{sec:test}
In Sec.~\ref{sec:result}, we found that some events can be explained by 
both the delayed wind and delayed jet models. Therefore, it is challenging 
to distinguish between these scenarios solely based on radio observations.
In this section, we investigate whether multiwavelength observations using 
TeV gamma-ray (CTAO), x-ray ({\it Chandra}), optical (Rubin, ZTF), 
and radio (ALMA, SKAO, VLA, MeerKAT) facilities can serve as diagnostics to discriminate among the instantaneous wind, delayed wind, and delayed jet models.  

We compute both the synchrotron and SSC emission for each model using the following parameter sets. For the instantaneous wind model, we adopt 
${\mathcal E}_{k}^{\rm iso}=5.0 \times 10^{51}\,{\rm erg}$, 
and $v_0 = 0.1c$.  
For the delayed wind model, we use 
${\mathcal E}_{k}^{\rm iso}=5.0 \times 10^{52}\,{\rm erg}$, 
and $v_0 = 0.1c$.  
For the delayed jet model, we adopt 
$\theta_v = 0$, $\theta_0 = 0.35\,{\rm rad}$, 
${\mathcal E}_{k}^{\rm iso}=5.0 \times 10^{52}\,{\rm erg}$, 
$\Gamma_0 = 3$, and $T_{\rm del}=1.0\times10^8\,{\rm s}$.
In both the delayed wind and delayed jet models, we consider
$T_{\rm del}=1.0\times10^7\,{\rm s}$ and $1.0\times10^8\,{\rm s}$.
In all cases, we fix the external medium and microphysical parameters to 
$n_{\rm ext} = 10^3\,{\rm cm^{-3}}$, $w = 1.0$, $s = 2.4$, 
$\epsilon_B = 0.01$, $\epsilon_e = 0.1$, and $f_e = 0.1$.  
We assume a redshift of $z = 0.014$ ($d_L = 60~{\rm Mpc}$), representative of nearby optical TDEs such as iPTF16fnl and more recent events including AT~2025aarm \citep{Miller2025}.

\begin{figure*}
\centering
\begin{minipage}{0.45\linewidth}
\centering
\includegraphics[width=1.0\textwidth]{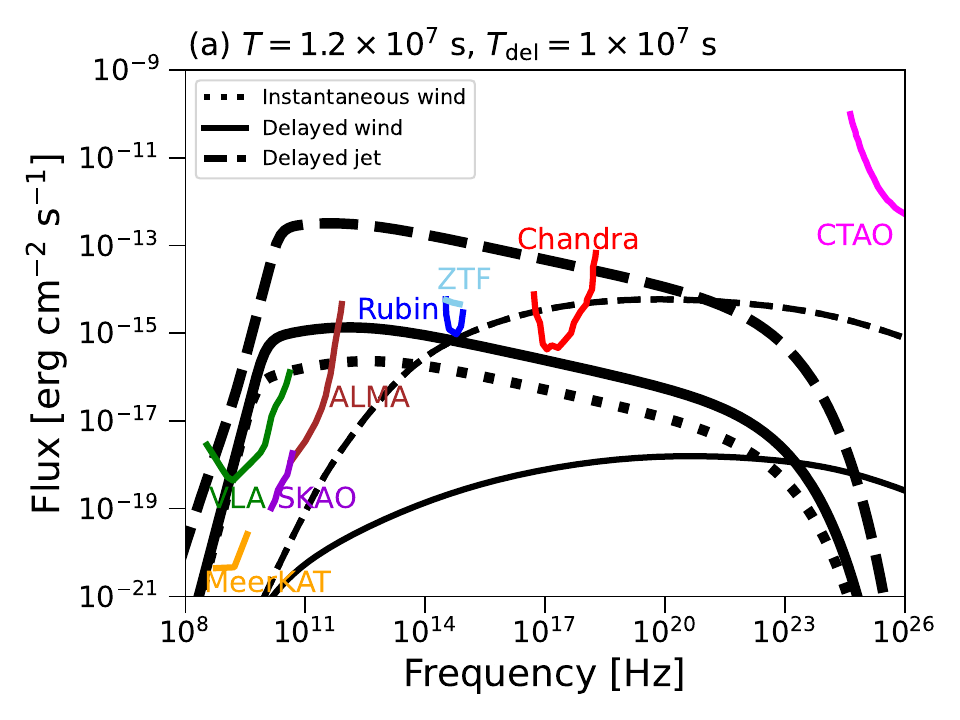}
\end{minipage}
\begin{minipage}{0.45\linewidth}
\centering
\includegraphics[width=1.0\textwidth]{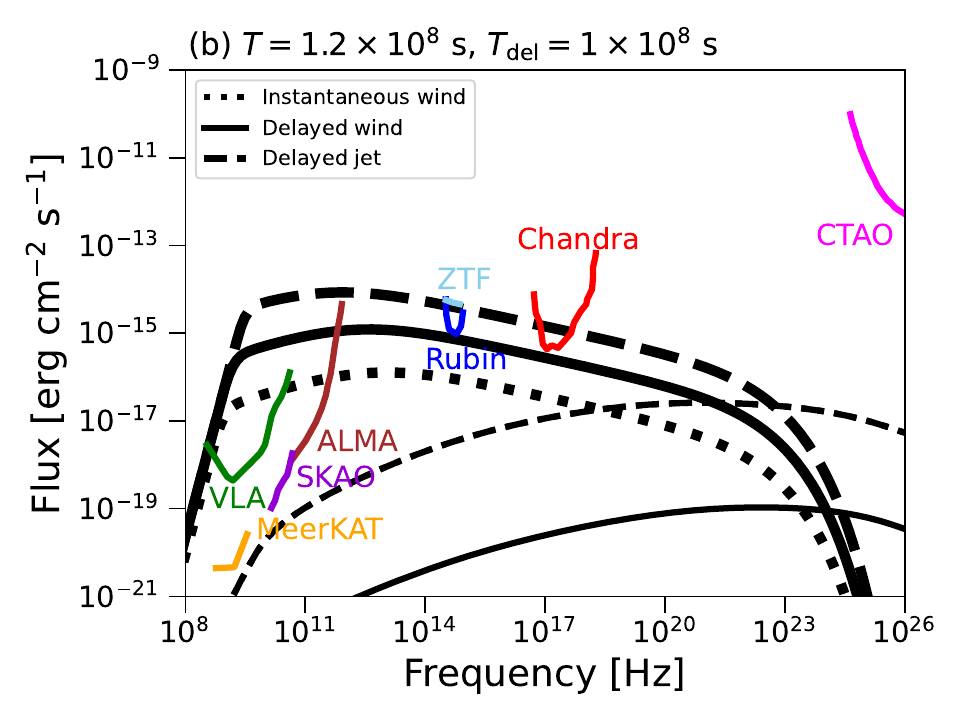}
\end{minipage}
\vspace{-0.3cm}
\caption{
Figure~\ref{fig:multi} shows predicted multiwavelength emission
from the instantaneous wind, delayed wind, and delayed
jet models at different observer times measured from the launch of the instantaneous outflow ($T$) and delay times ($T_{\rm del}$) are made:
$T=1.2\times10^7\,{\rm s}$ and $T_{\rm del}=1.0\times10^7\,{\rm s}$ in panel~(a), and
$T=1.2\times10^8\,{\rm s}$ and $T_{\rm del}=1.0\times10^8\,{\rm s}$ in panel~(b).
The black dotted, solid, and dashed curves correspond to the 
instantaneous wind, delayed wind, and delayed jet models, respectively.
Synchrotron and SSC components are shown as thick and thin curves, respectively.
The SSC component is calculated for all models but is negligible for the instantaneous wind model.
Colored curves represent the sensitivity limits of various detectors
gamma-ray (magenta: CTAO~\citep{CTA}, 50 h), 
x-ray (red: Chandra~\citep{Chandra}, 100 ks),
optical (blue: Rubin~\citep{Rubin}, 30 s; sky-blue: ZTF~\citep{ZTF}, 30 s),
and radio 
(brown: ALMA~\citep{ALMA}, 1 h; 
violet: SKAO~\citep{SKA}, 10 h; 
green: VLA~\citep{VLA}, 1 h; 
orange: MeerKAT~\citep{MeerKAT}, 1 h). 
Considering AT 2025aarm as an example, a source distance of $z = 0.014$ ($d_L = 60~{\rm Mpc}$) is assumed. Only the delayed jet model exceeds the x-ray and optical sensitivity limits at these epochs for a nearby TDE.
}
\label{fig:multi}
\end{figure*}

\begin{figure*}
\centering
\begin{minipage}{0.45\linewidth}
\centering
\includegraphics[width=1.0\textwidth]{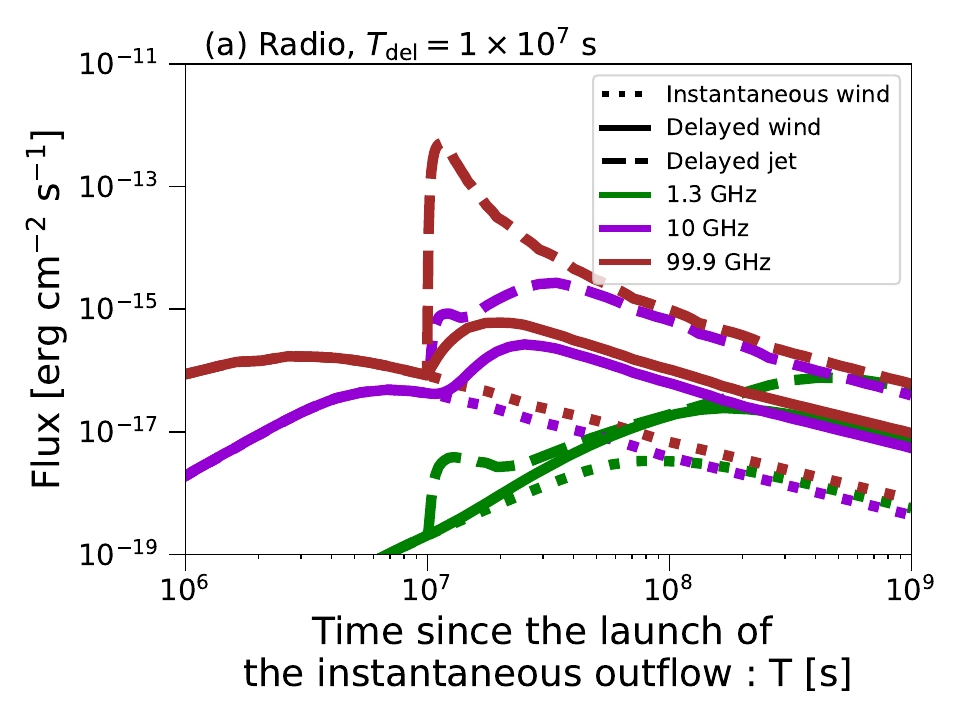}
\end{minipage}
\begin{minipage}{0.45\linewidth}
\centering
\includegraphics[width=1.0\textwidth]{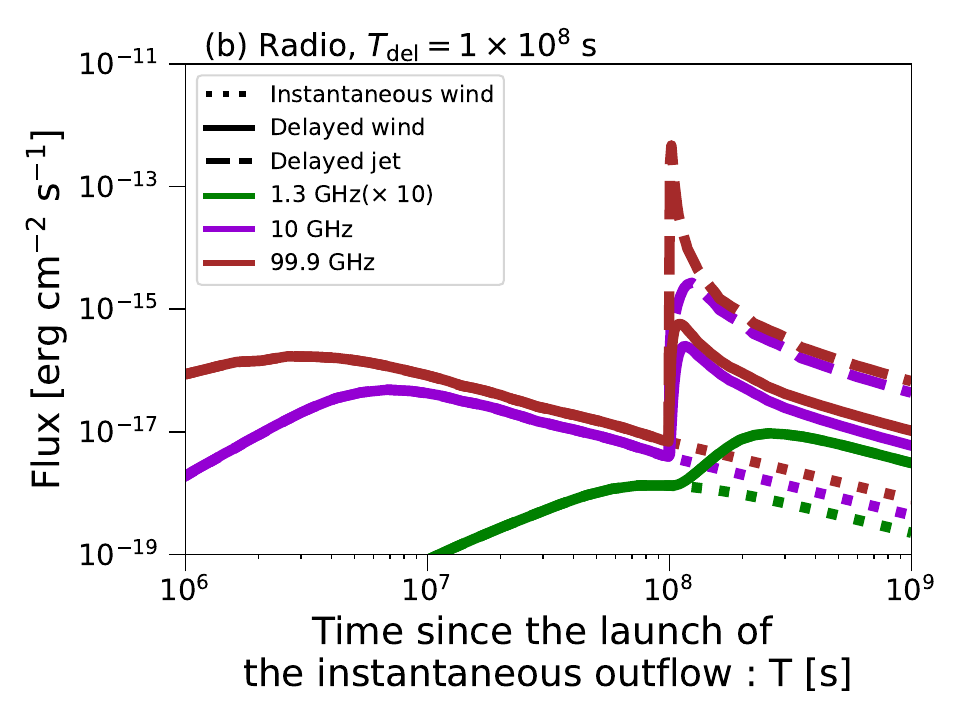}
\end{minipage}
\begin{minipage}{0.45\linewidth}
\centering
\includegraphics[width=1.0\textwidth]{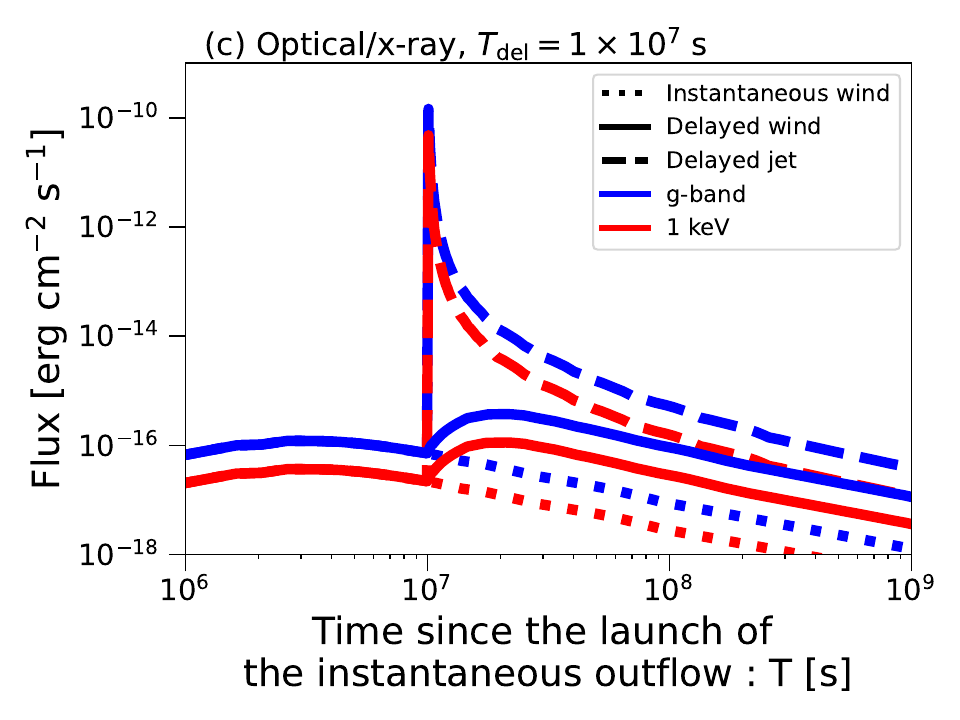}
\end{minipage}
\begin{minipage}{0.45\linewidth}
\centering
\includegraphics[width=1.0\textwidth]{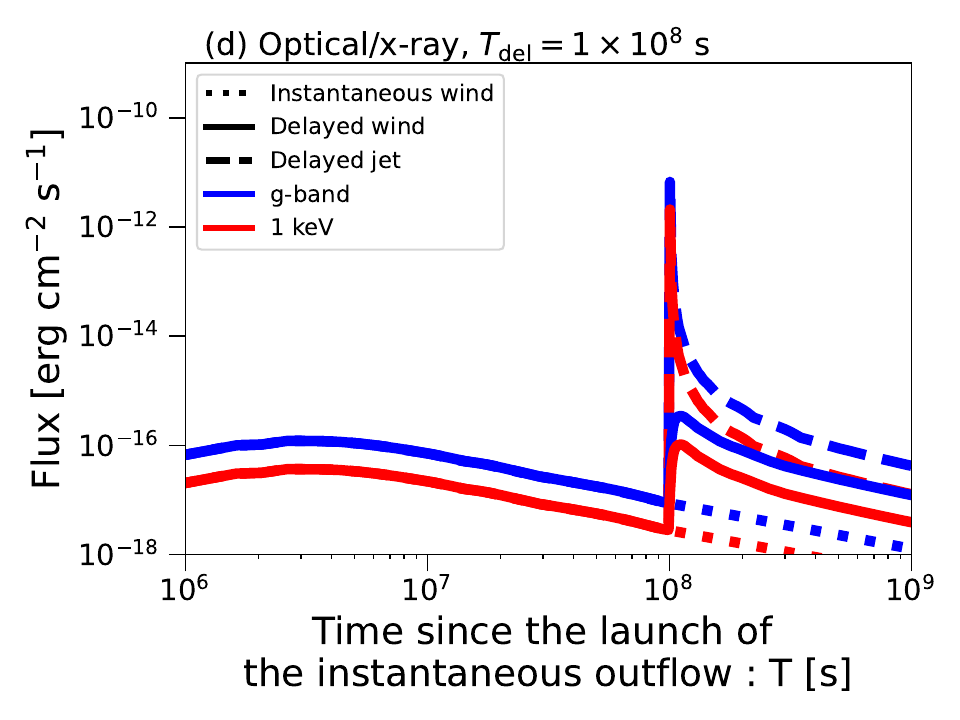}
\end{minipage}
\begin{minipage}{0.45\linewidth}
\centering
\includegraphics[width=1.0\textwidth]{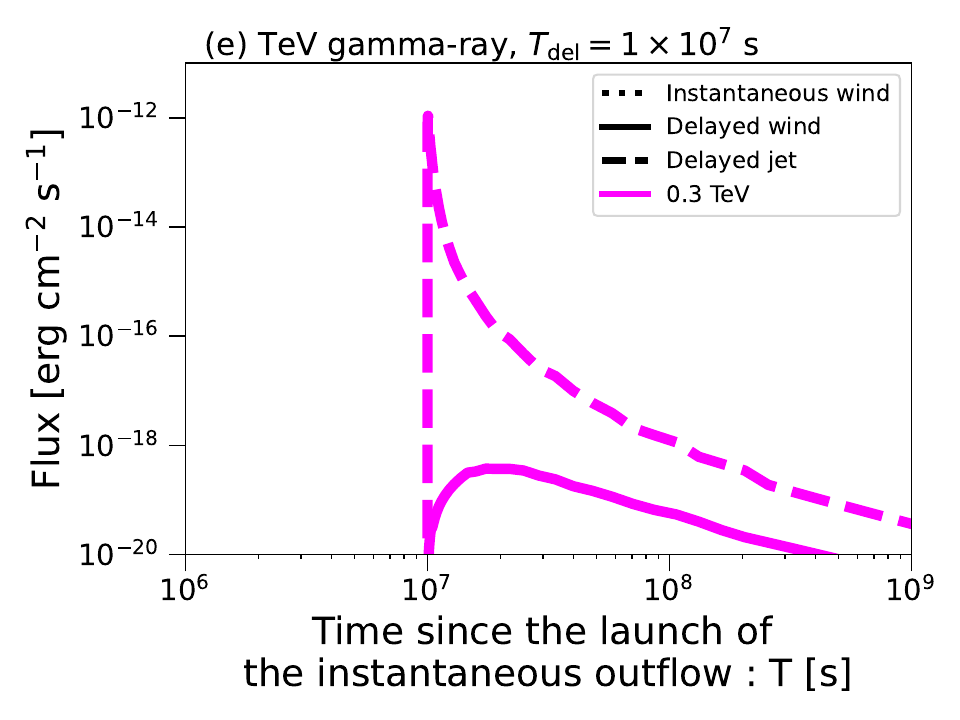}
\end{minipage}
\begin{minipage}{0.45\linewidth}
\centering
\includegraphics[width=1.0\textwidth]{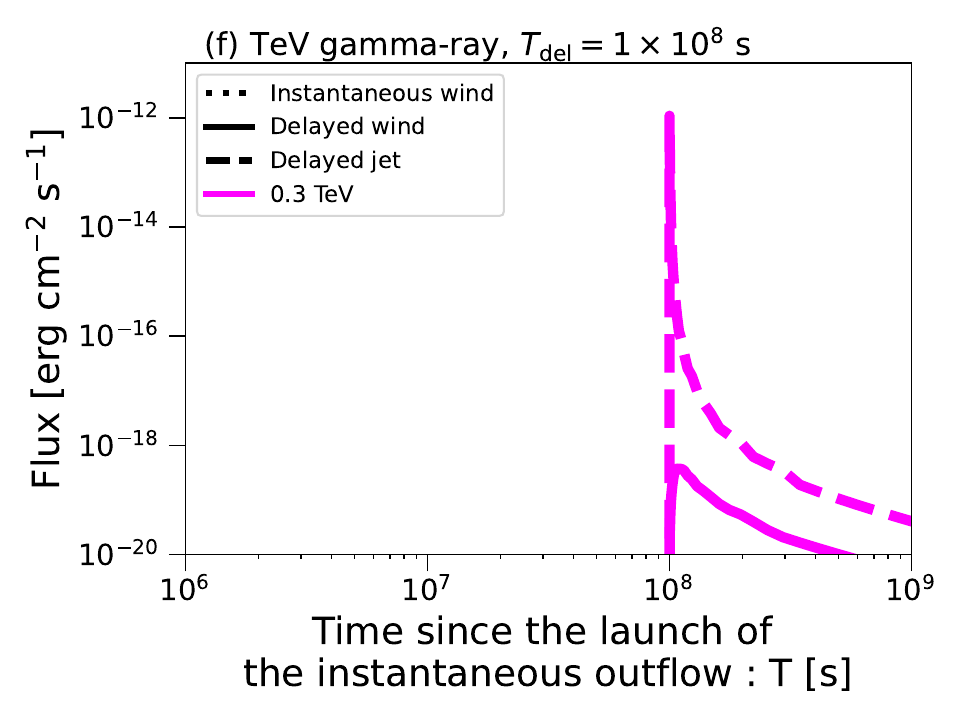}
\end{minipage}
\vspace{-0.3cm}
\caption{ 
Figure~\ref{fig:multi_lc} presents the multiwavelength light curves in the radio (panels (a) and (b)), optical/x-ray (panels (c) and (d)), and TeV gamma-ray (panels (e) and (f)) bands for the same model configurations and parameter sets adopted in Fig.~\ref{fig:multi}. Panels (a), (c), and (e) correspond to $T_{\rm del}=1\times10^7$~s, whereas panels (b), (d), and (f) correspond to $T_{\rm del}=1\times10^8$~s.
}
\label{fig:multi_lc}
\end{figure*}

We present the results in Figs.~\ref{fig:multi} and \ref{fig:multi_lc}.
The x-ray telescope {\it Chandra} has the potential to detect the synchrotron emission from the delayed jet model regardless of the adopted delay time (see dashed lines in Fig.~\ref{fig:multi}). In contrast, the x-ray from the instantaneous wind and delayed wind models remains challenging to detect with {\it Chandra} because the corresponding outflows are nonrelativistic and therefore lack strong Doppler boosting (see solid and dotted lines in Fig.~\ref{fig:multi}).
As shown in Fig.~\ref{fig:multi_lc}, the delayed jet emission evolves rapidly after launch, implying that the x-ray counterpart is detectable only near the launching epoch of the delayed outflow. 
Since the delay time itself is long ($T_{\rm del}\sim10^{7-8}\,{\rm s}$), this emission would appear observationally as late-time x-ray brightening, providing an important signature of delayed outflow activity.

Interestingly, recent {\it Chandra} observations of the nearby TDE AT~2025aarm reported an x-ray flux of $\sim10^{-13}\ {\rm erg\ cm^{-2}\ s^{-1}}$ at $T\sim10^7\,{\rm s}$ \citep{Gaurava2026}. This flux level is comparable to the synchrotron x-ray emission predicted by our delayed jet model in Fig.~\ref{fig:multi}(a). Late-time x-ray observations can provide a powerful probe of delayed relativistic outflows in nearby TDEs. 
Therefore, late-time x-ray detections provide an important diagnostic for distinguishing relativistic delayed jets from nonrelativistic delayed winds, even when the radio light curves are similar.

In the optical band, the Rubin Observatory has the ability to detect synchrotron emission from the delayed jet model for both short and long delay times considered here (see dashed lines in Fig.~\ref{fig:multi}). On the other hand, detection with ZTF may be more challenging because of its lower sensitivity, although even close events may still become detectable. The detectability at different source distances is discussed further in Sec.~\ref{subsec:multi}. 
However, we note that identifying a nonthermal synchrotron component may be complicated by the presence of bright thermal optical emission from the TDE accretion flow itself.
Nevertheless, the thermal optical emission is expected to exhibit a quasi blackbody spectrum, whereas the synchrotron component follows a nonthermal power-law spectrum. Multiband photometry and/or spectroscopy may therefore help disentangle the delayed synchrotron emission from the thermal TDE component.

TeV gamma-ray emission would be difficult to detect with CTAO because the SSC component remains below the CTAO sensitivity for typical TDE distances. 
For example, TeV follow-up observations have been reported for the nearby TDE AT~2025aarm \citep{CTAO2025}.  Although the SSC component predicted by our model remains challenging to detect, TeV emission may become detectable for smaller values of $\epsilon_B$ than assumed in our fiducial model, which would lead to a larger Compton-$Y$ parameter and hence stronger SSC emission. 
Furthermore, additional inverse-Compton components, such as external inverse-Compton (EIC) emission, may further enhance the TeV gamma-ray flux, as discussed in Sec.~\ref{subsec:multi}.

At radio frequencies, all three models are potentially detectable at low frequencies, such as those observed with MeerKAT or VLA. 
On the other hand, the emission from the delayed jet model exceeds the sensitivity threshold in the high frequency regime accessible by ALMA. 
Although ALMA's sensitivity decreases toward higher frequencies, the delayed jet remains in a transrelativistic phase at $\sim 10^8$~s. 
In contrast, both the instantaneous and delayed wind models have already entered the nonrelativistic phase by that time, resulting in significantly lower fluxes. 
As a result, the delayed jet produces the highest flux among the models at these high frequencies, making it detectable by ALMA, despite its more limited sensitivity in this regime.


\section{Discussion}
\label{sec:discussion}

\begin{figure*}
\centering
\begin{minipage}{0.45\linewidth}
\centering
\includegraphics[width=1.0\textwidth]{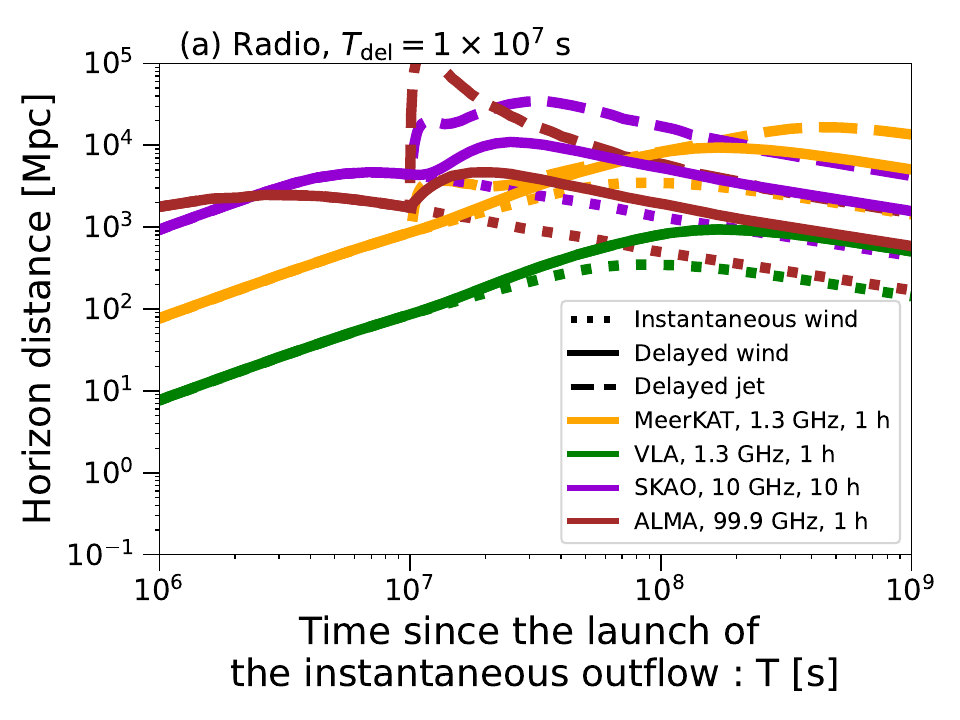}
\end{minipage}
\begin{minipage}{0.45\linewidth}
\centering
\includegraphics[width=1.0\textwidth]{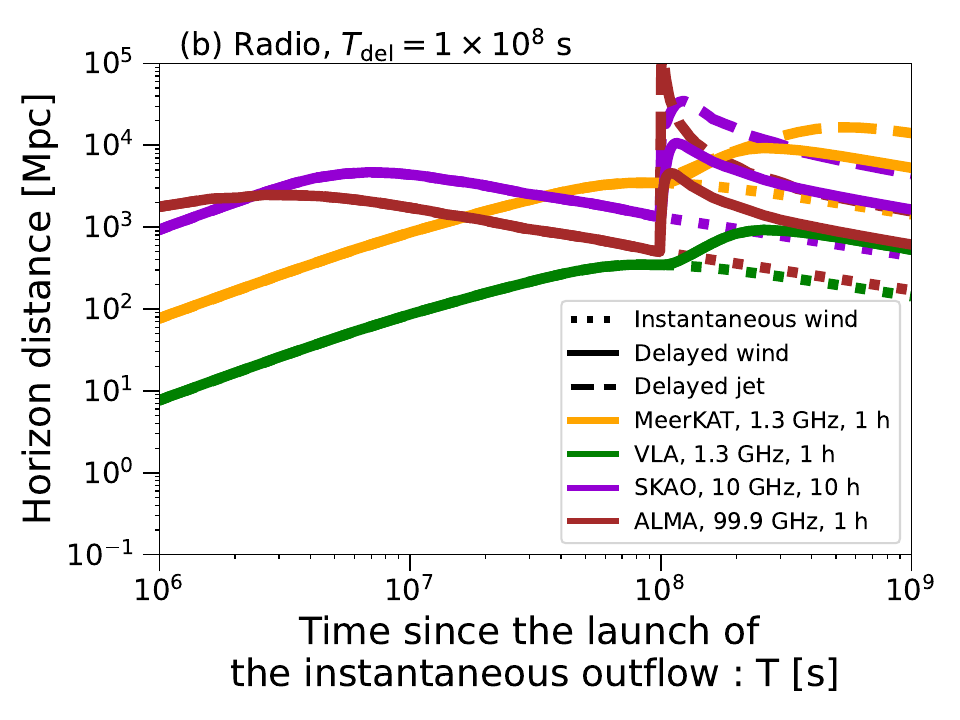}
\end{minipage}
\begin{minipage}{0.45\linewidth}
\centering
\includegraphics[width=1.0\textwidth]{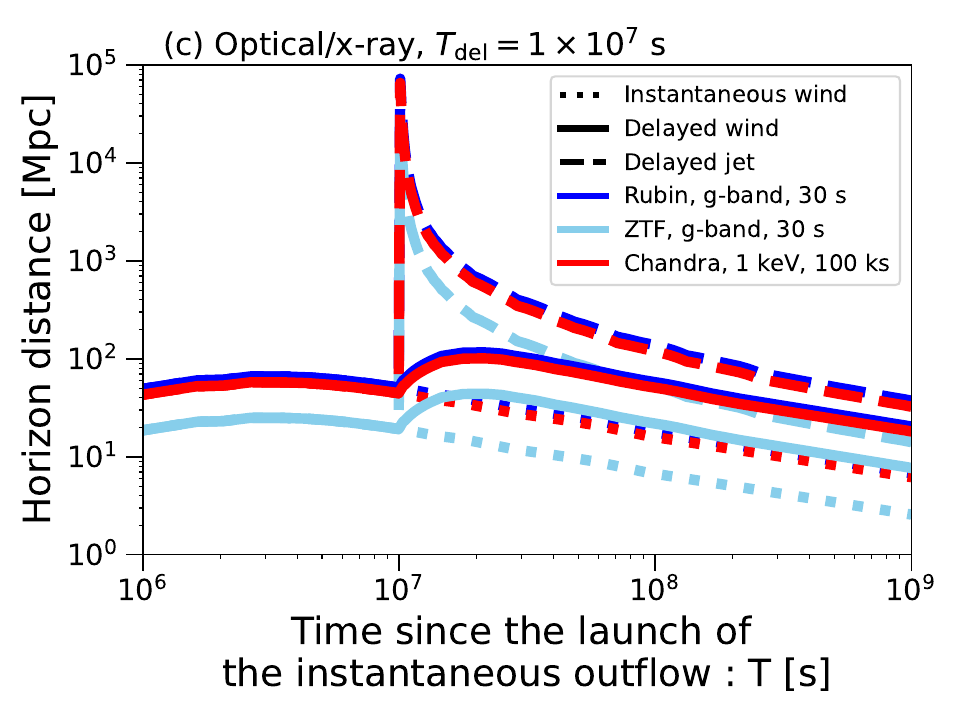}
\end{minipage}
\begin{minipage}{0.45\linewidth}
\centering
\includegraphics[width=1.0\textwidth]{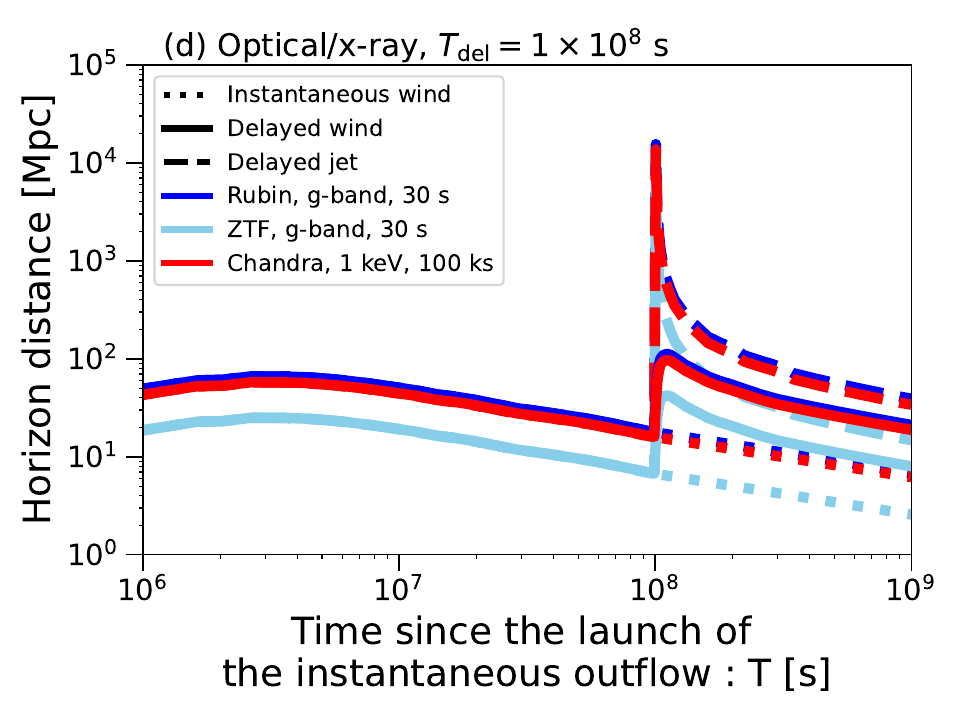}
\end{minipage}
\begin{minipage}{0.45\linewidth}
\centering
\includegraphics[width=1.0\textwidth]{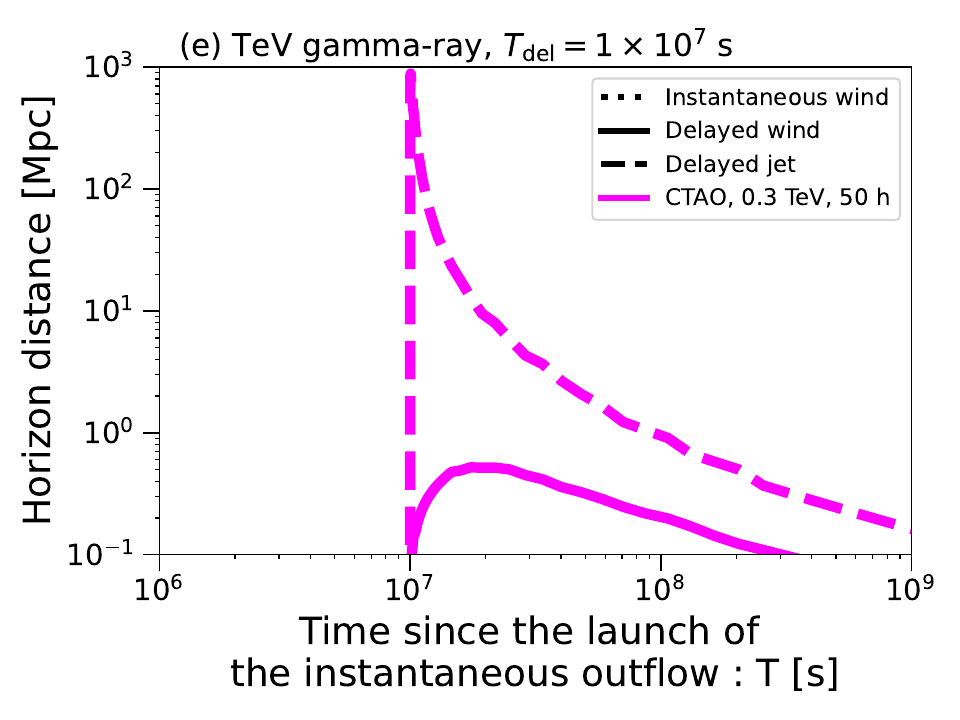}
\end{minipage}
\begin{minipage}{0.45\linewidth}
\centering
\includegraphics[width=1.0\textwidth]{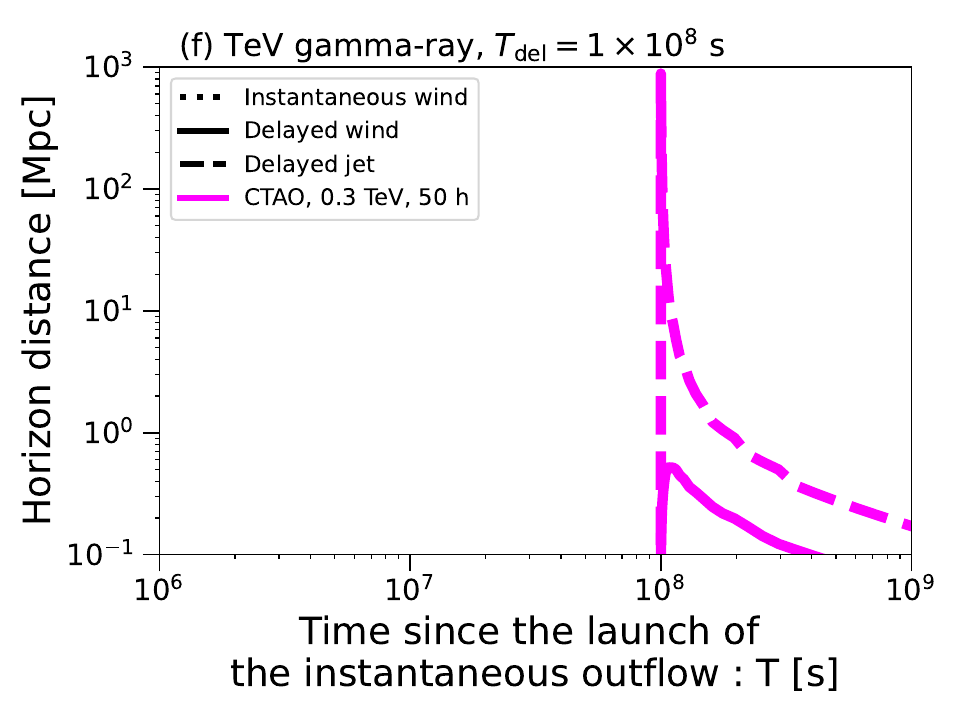}
\end{minipage}
\vspace{-0.3cm}
\caption{Figure~\ref{fig:horizon} shows the multiwavelength detection horizons in the radio (panels (a) and (b)), optical/x-ray (panels (c) and (d)), and TeV gamma-ray (panels (e) and (f)) bands for the same models and parameter sets in Fig.~\ref{fig:multi}. Panels (a), (c), and (e) correspond to $T_{\rm del}=1\times10^7$~s, while panels (b), (d), and (f) correspond to $T_{\rm del}=1\times10^8$~s.
The horizons for MeerKAT, VLA, SKAO, ALMA, Rubin, ZTF, and Chandra are set by synchrotron emission, whereas the CTAO horizon is set by SSC emission.
In the optical and x-ray bands, only the delayed jet model is detectable beyond $\sim 100$~Mpc, while both instantaneous wind and delayed wind models are limited to much smaller distances.
The multiwavelength facilities considered here are:
MeerKAT (orange, 1 h; panel a), VLA (green, 1 h; panel a),
SKAO (violet, 10 h; panel a), ALMA (brown, 1 h; panel a),
Rubin (blue, 30 s; panel b), ZTF (sky-blue, 30 s; panel b),
Chandra (red, 100 ks; panel b), and CTAO (magenta, 50 h; panel c). 
}
\label{fig:horizon}
\end{figure*}

\subsection{Energetics and delayed launching}

As shown in Table~\ref{tab:fit}, for the delayed radio flare TDEs, the isotropic kinetic energy inferred for the delayed wind and delayed jet components is an order of magnitude larger than that of the instantaneous wind.
This trend may reflect the fact that the late-time ($\gtrsim 10^8\,{\rm s}$) radio emission is brighter than the early time emission ($\lesssim 10^8\,{\rm s}$), requiring a more energetic outflow to explain the observed flux.
On the other hand, the physical mechanisms responsible for launching such powerful outflows at delayed times remain uncertain. 
Several possibilities have been proposed to explain the delayed launching of outflows in TDEs. These include a time lag in the formation of an accretion disk due to inefficient circularization of stellar debris {\citep{Guillochon:2015qfa,Hayasaki:2015pxa,Bonnerot:2015ara} or the disk transition due to the change of the accretion rate \citep{Murase:2020lnu}, delayed jet launch resulting from the gradual accumulation of magnetic flux and the eventual activation of the Blandford–Znajek mechanism \citep{Piran:2015ija}, and episodic reactivation of accretion driven outflows by fallback rate fluctuations \citep{vanVelzen:2020cwu}.
Such delayed processes may allow the subsequent production of more massive and energetic outflows, than what would be expected from prompt ejection.
In contrast, the TDE iPTF16fnl, which does not exhibit a delayed radio flare, shows an isotropic kinetic energy for the delayed wind component that is comparable to that inferred for the instantaneous wind.
This is because iPTF16fnl has a peak radio luminosity of $\sim10^{37}\,{\rm erg\;s^{-1}}$, which is about two orders of magnitude dimmer than that of the delayed radio flare TDEs. This suggests that outflows launched promptly after disruption may be intrinsically less energetic.

Radio rebrightening events in our sample appear to require the presence of two distinct outflow components, either in the form of two winds or a combination of a wind and a jet (see Table~\ref{tab:fit}).
Nevertheless, even for TDEs that exhibit pronounced late-time radio flares, such as AT~2019ehz and AT~2019eve, the inclusion of a delayed outflow component is not strictly necessary to reproduce the observed light curves (see Table~\ref{tab:fit} and Fig.~\ref{fig:result3}).
This diversity in radio behavior likely reflects underlying differences in the physical properties of the accretion disk, such as the efficiency of debris circularization, the timescale for disk formation, or the ability to accumulate magnetic flux, which may in turn regulate whether additional outflows are produced and at what epoch they are launched.

The fallback time of the most tightly bound debris can be estimated as
$T_{\rm fb} \sim 10^{6}\,{\rm s}~
(M_{\rm BH}/10^{7}\,M_\odot)^{1/2}
(M_*/M_\odot)^{-1}
(R_*/R_\odot)^{3/2}$,
where $M_{\rm BH}$ is the BH mass and $M_*$/$R_*$ denote the mass/radius of the disrupted star, respectively~\citep{Rees1988}.
By contrast, the inferred delay times for launching outflows are $T_{\rm del} \sim 10^{7-8}\,{\rm s}$ (see Table~\ref{tab:parameter}), which are at least an order of magnitude longer than $T_{\rm fb}$. 
Since the mass fallback rate monotonically declines as $\dot{M}_{\rm fb} \propto T^{-5/3}$ for $T \gtrsim T_{\rm fb}$~\citep{Rees1988}, fallback driven accretion alone cannot account for the observed late-time radio brightening. 
This discrepancy motivates the consideration of additional delayed processes capable of powering energetic outflows at late times, such as inefficient debris circularization, gradual accumulation of magnetic flux leading to the activation of a Blandford–Znajek jet, or episodic reactivation of accretion triggered by fluctuations in the fallback rate.

A variety of physical mechanisms have been proposed to account for the delayed launching of outflows in TDEs. 
One possibility is a time lag in accretion disk formation caused by inefficient debris circularization or long viscous timescales~\citep{Murase:2020lnu,Hayasaki:2021ggg,Cendes2023}.
Late-time radio activity observed in a significant fraction of optically selected TDEs may support such delayed accretion phases~\citep{Alexander:2025wnh}. 
Delayed mass ejection may also arise from disk instabilities or from the gradual accumulation of magnetic flux that eventually triggers the Blandford–Znajek process~\citep{Wu2025,Kelley2014,Tchek2014}. 
In addition to central engine physics, environmental and geometric effects can delay the observability of outflows; for example, interaction with a structured CNM or off-axis viewing may postpone the emergence of detectable synchrotron emission~\citep{NG2007,Giannios2011,Mimica2015,Sato2024}.
Overall, the delayed radio behavior of TDEs likely reflects a combination of accretion physics, magnetic field evolution, and circumnuclear environment.

Based on our modeling, we find that the observed early and delayed radio emission properties of all events can be consistently reproduced with isotropic-equivalent kinetic energies clustered around ${\mathcal E}_{k}^{\rm iso}\sim5\times10^{51}$–$5\times10^{52}\,\mathrm{erg}$ (see Table~\ref{tab:parameter}).
This clustering suggests that winds and jets in optical TDEs may share a comparable energy budget. These values may be systematically larger than those inferred in previous works (e.g., Ref.~\citep{Cendes2023}), where the kinetic energy was estimated under the assumption of equipartition between relativistic electrons and magnetic fields. 
Our calculation follows the time-dependent hydrodynamical evolution of winds and jets and constrains the microphysical parameters ($\epsilon_e$, $\epsilon_B$, and $f_e$) through radio fitting rather than assuming equipartition. 
Consequently, differences between our estimates and the equipartition based estimates of Ref.~\citep{Cendes2023} are expected.

To demonstrate this more explicitly, we consider the delayed wind model for ASASSN-14ae. 
At the epochs considered below, the delayed wind has already entered the deceleration phase but remains outside the deep-Newtonian regime.
The second 5 GHz data point at $T\approx2696 \,{\rm d}$ may correspond to the epoch when the synchrotron self-absorption frequency crosses the observing band, $\nu_a\approx5\,{\rm GHz}$. 
Using Eq.~(\ref{nu_a}), the condition $\nu_a(2696\,{\rm d})\approx5\,{\rm GHz}$ implies $n_0^{0.3}E_{52}^{0.35}\approx15$.
Prior to this epoch, the 5 GHz emission lies in the regime $\nu_m<\nu<\nu_a$.
Evaluating Eq.~(\ref{Fnu_degen}) at $T=3243\,{\rm d}$ with the measured flux $F_{\nu=5\,{\rm GHz}}\approx0.4\,{\rm mJy}$ yields $n_0^{-11/16}E_{52}^{7/16}\sim0.01$.
Solving these relations with representative microphysical parameters and delay time inferred from our numerical modeling ($s$, $w$, $\epsilon_e$, $\epsilon_B$, $f_e$, and $T_{\rm del}$) gives
${\mathcal E}_{k}^{\rm iso}\sim{\rm few}\times10^{52}\,{\rm erg}$.
While this estimate is not fully independent of the detailed modeling, it provides an analytic consistency check that the observed flux level and self-absorption crossing time require large kinetic energies.

Importantly, inferred kinetic energies of ${\mathcal E}_{k}^{\rm iso}\sim10^{52}\,\mathrm{erg}$ may lie within the available energy budget for mass accretion onto a SMBH, which can be physically realized in delayed outflow scenarios. The formation of an accretion disk can be delayed due to the inefficient circularization of the stellar debris \citep{Guillochon:2015qfa,Hayasaki:2015pxa,Bonnerot:2015ara}, where the circularization and angular-momentum transport are inefficient, such that a fraction of the bound debris is promptly accreted, while a substantial fraction of the debris remains on extended orbits and accretes at later times. 
Such delayed accretion, involving a sizable fraction of the stellar mass, is sufficient to account for the inferred outflow energetics, and the gravitational energy budget is estimated to be
\footnotesize
\begin{eqnarray}
{\mathcal E}_{\rm grav}&=&\eta_{\rm acc}f_{\rm in}f_{\rm fb}M_*c^2\nonumber\\
&\simeq&0.9\times10^{52}~\left(\frac{\eta_{\rm acc}}{0.1}\right)\left(\frac{f_{\rm in}}{0.1}\right)\left(\frac{f_{\rm fb}}{0.5}\right)\left(\frac{M_*}{1~M_\odot}\right)\,\mathrm{erg},
\end{eqnarray}
\normalsize
where $M_*$ is the stellar mass, $f_{\rm in}\sim0.1$ is the mass fraction of accreting material, and $f_{\rm fb}\sim0.5$ is the fraction of the stellar mass that falls back (e.g., Ref.~\citep{Murase:2020lnu}).
This indicates that the required outflow energy remains within the available energy budget.

Alternatively, delayed outflows may be associated with the disk transition (e.g., Ref.~\citep{Murase:2020lnu}). Delayed outflows may be launched when the luminosity declines to a critical value of $L_{\rm crit} \sim 0.02\,L_{\rm Edd}$ \citep{Goodwin2026}. The total energy available from such an outflow over the delay timescale is therefore $\sim L_{\rm crit}T_{\rm del}\sim 3\times10^{51}(M_{\rm BH}/10^7\,M_\odot)(T_{\rm del}/10^8\,\mathrm{s})\,\mathrm{erg}$. This may account for the collimation-corrected kinetic energy in the delayed jet model, ${\mathcal E}_{k}^{\rm jet} \sim {\mathcal E}_{k}^{\rm iso}\theta_0^2/2 \sim 3\times10^{51}\,\mathrm{erg}$. For a more massive BH ($M_{\rm BH}\sim10^8\,M_\odot$), the available energy increases to $\sim 3\times10^{52}(M_{\rm BH}/10^8\,M_\odot)(T_{\rm del}/10^8\,\mathrm{s})\,\mathrm{erg}$, which is comparable to the isotropic kinetic energy adopted in the delayed wind scenario. At such large BH masses, ordinary main-sequence stars are close to the direct-capture limit, whereas evolved stars with extended envelopes can still be disrupted outside the horizon. Hence, a TDE around
$M_{\rm BH}\sim10^8M_\odot$ may more involve an evolved star such as
a red giant. In giant-star disruptions, the extended envelope generally leads to
a longer fallback time and lower fallback rate, often sub-Eddington, although
short super-Eddington episodes may occur depending on the core-envelope
structure and encounter depth (e.g., Ref.~\citep{Gezari:2021bmb,MacLeod2012}). 
If this is the case, our results imply that delayed radio-emitting outflows preferentially originate from TDEs hosting massive BHs with $M_{\rm BH}\sim10^8~M_\odot$.

\subsection{Implications for multiwavelength observations}
\label{subsec:multi}

In the delayed jet scenario, a relativistic outflow is launched at $T_{\rm del} = 10^8\,{\rm s}$ after the stellar disruption. 
Our numerical calculations show that the synchrotron emission from this jet peaks around this epoch, with fluxes in both x-ray and optical bands exceeding the sensitivity limits of \emph{Chandra} and Rubin, respectively. Notably, the optical emission remains above Rubin’s detection threshold for a narrow time window of $\approx (1.0$--$1.2) \times 10^8\,{\rm s}$. After this, the flux declines rapidly due to jet deceleration and geometric spreading, quickly falling below the detectability threshold.
With the commencement of full scale operations at the Rubin Observatory, a substantial increase in the detection rate of optical transients is anticipated.
Consequently, it is essential to establish optimal timing for follow-up observations using x-ray and radio facilities subsequent to the identification of optical emission by Rubin. 
Once a rising optical signal is detected around $T \sim 10^8~{\rm s}$ after the launch of the instantaneous outflow, follow-up x-ray observations with \emph{Chandra} should be initiated within approximately $10^6$–$10^7\,{\rm s}$. This timescale ensures that the transient x-ray counterpart is captured before it declines below the instrument’s sensitivity, as the x-ray light curve in the delayed jet scenario typically peaks and declines on timescales of weeks to a few months after the jet is launched.

Although the overall delay in jet or wind launching is on the order of years ($\sim 10^8\,{\rm s}$), once the outflow is initiated the subsequent emission evolves on much shorter timescales ($\sim 10^6$–$10^7\,{\rm s}$) as seen in Fig.~\ref{fig:multi_lc}. 
In particular, the rise and peak of the delayed component can develop rapidly. 
Therefore, high cadence optical monitoring around $T \sim 10^8\,{\rm s}$ is essential for discriminating between nonrelativistic winds and relativistic jets. 
A cadence of $\sim 10^6\,{\rm s}$ is recommended during this critical phase to resolve the rise and peak of the delayed emission with sufficient temporal resolution.
In contrast, radio emission remains detectable with ALMA, SKAO, VLA, and MeerKAT throughout the evolution in all models, namely the instantaneous wind, delayed wind, and delayed jet, owing to the lower absorption frequencies and the slower flux decay.
The rise and peak of the radio light curve typically occur when the absorption frequency $\nu_a$ crosses the observing band, developing over timescales of weeks to months. 
To capture this evolution effectively, radio follow-up should begin promptly after the detection of late-time optical emission (within a few days), with a recommended cadence of $\sim 10^6\,{\rm s}$.

We comment on how $T_{\rm del}$ should be defined. A practical approach is to identify the onset of rebrightening in the radio light curve, which marks the interaction of a newly launched outflow with the CNM. In this case, $T_{\rm del}$ is inferred as the epoch when the radio flux begins to rise after an initial decline. For an initial nonrelativistic wind outflow, nonthermal optical or x-ray counterparts are not expected. A more direct constraint could be obtained if nonthermal optical or x-ray emission is detected simultaneously with the radio rebrightening, as predicted in delayed jet scenarios.
In such cases, the rise of the high energy component would provide a more accurate measurement of $T_{\rm del}$.
Therefore, systematic multiwavelength follow-up is essential.

We compute the detection horizons for the telescopes considered in Sec.~\ref{sec:test} using Eq.~(8) of Ref.~\citep{Yuan:2021jjt}.  
The model parameters are the same as those adopted in Sec.\ref{sec:test}.
As shown in Figs.~\ref{fig:horizon}(c) and (d), both {\it Chandra} and Rubin are capable of detecting the delayed jet emission when the source is located within $\approx 10^5$~Mpc near the peak epoch. However, because the delayed jet emission rapidly declines after the peak, such a large horizon is maintained only for a short duration. In practice, the effective detection distance is expected to be much smaller, likely limited to several hundred Mpc for typical follow-up observations.
The optical and x-ray emission from the delayed wind model is detectable out to $\approx 100$~Mpc, whereas the instantaneous wind model can be detected within $\approx 20$~Mpc due to its nonrelativistic nature and the absence of significant Doppler boosting.
Rubin and ZTF operate in the same optical band, but their sensitivities are significantly different (see blue and sky-blue lines in Fig.~\ref{fig:multi}), resulting in different horizon distances. 
The overlap between Rubin and Chandra reflects their similar detection horizons.
Our results further indicate that CTAO can detect the TeV gamma-ray emission if the source is nearby: within $d_L \lesssim 0.4$~Mpc for the delayed wind model and $d_L \lesssim 10^3$~Mpc for the delayed jet model (see Figs.~\ref{fig:horizon}(e) and (f)).  
However, because the delayed jet emission declines rapidly after the peak, the TeV detection horizon of $\sim10^3$~Mpc applies only during a short lived phase. In practice, the effective horizon for CTAO observations is expected to be significantly smaller, likely limited to a few Mpc.

Radio facilities such as SKAO and ALMA have substantially larger detection horizons (see violet and brown lines in Figs.~\ref{fig:horizon}(a) and (b)). 
Because the delayed components become radio bright at late times, these instruments can detect the emission even if the source lies at cosmological distances up to $\sim 10$~Gpc at the time the delayed outflows (jet and/or wind) are launched.  
This implies that radio transients that brighten over several years and fade on similarly long timescales, with little or no accompanying x-ray or optical counterparts, may plausibly originate from delayed jets and/or winds associated with a TDE. Such scenarios highlight the importance of wide-field, high-sensitivity radio surveys in uncovering the otherwise hidden population of delayed TDE components.

In Sec.~\ref{sec:test}, we investigated the SSC emission.
For the delayed wind and delayed jet models, two distinct outflows are present, which provide two different photon fields. 
This raises the possibility of EIC scattering, in which electrons in the instantaneous wind shock upscatter photons originating from the delayed component.  
In both models, the energy density of the seed photons from the delayed component is far below the magnetic energy density in the instantaneous wind shock.  
Another possible source of seed photons is the thermal emission from the disk \citep{Lu2016}. 
Such thermal EIC emission can become important at early times and small radii, where the external radiation field is sufficiently intense. 
In our scenario, the emission considered here mainly arises from the external forward shock at relatively large radii, where the thermal photon energy density becomes smaller than the post-shock magnetic energy density. 
Therefore, thermal EIC emission is expected to be subdominant for the late-time external forward shock considered here. 
However, EIC emission may become significant in other dissipation regimes occurring at smaller radii \citep{Lu2016}.

\subsection{Comparison with other recent events and implications for population studies}

In addition to the events analyzed in this work, recent observations suggest the presence of delayed outflow components in other TDEs as well~\citep{Sfaradi2025,Golay2025,Lin2025}. 
For instance, AT~2024tvd exhibits a rapid late-time rebrightening in the radio band, which appears to require an additional delayed component to explain the observed flux evolution~\citep{Sfaradi2025}. 
Similarly, WTP14adeqka does not show a clear rebrightening in the radio light curve,
but its detection at very late epochs with relatively high radio luminosity also indicates the presence of a delayed outflow~\citep{Golay2025}. 
In both cases, however, whether the delayed component is best explained by a nonrelativistic wind or a relativistic jet remains unclear and will require detailed modeling. 
These recent examples highlight the growing evidence that delayed outflows may be more common in TDEs than previously expected, and reinforce the importance of systematic multiwavelength follow-up efforts to characterize their nature.

On the other hand, the radio emission of AT~2018hyz also exhibits a late-time flare \citep{Cendes2022, Anumarlapudi2024}. 
Its radio light curve is characterized by a rapid rise between $\sim8 \times 10^7$~s and $\sim2 \times 10^8$~s. Such a long term rise in radio flux is not seen in delayed emission components, making it challenging for delayed outflow models to explain the behavior of AT~2018hyz. 
Alternative scenarios, such as an off-axis jet model, have been proposed to account for such emission~\citep{Sfaradi2023, Sato2024}. AT~2018hyz may also have two jet components~(see, e.g., Ref.~\citep{Sato2024}).

Note that jet-loud TDEs are rare events. The apparent rate density of on-axis jetted TDEs is $\sim 0.03~{\rm Gpc^{-3}\,yr^{-1}}$ (e.g., Ref.~\citep{Sun2015}). 
Correcting for beaming implies a true jetted TDE rate of $\sim 6~{\rm Gpc^{-3}~yr^{-1}}$, which is still smaller than that of jet-quiet TDEs ($\sim10^3~{\rm Gpc^{-3}\,yr^{-1}}$) (e.g., Refs.~\citep{Sun2015,Andreoni2022}). 
Delayed jets are expected to be launched several years after the initial disruption takes place and may not have been captured by previous radio surveys, implying that the currently inferred jetted TDE rate could be underestimated. 
Future systematic multiwavelength surveys will be essential to constrain the true occurrence rate of the delayed jet phenomena from TDEs.

\subsection{Model uncertainties and limitations}

For now, due to the observational limitations, several of the fitted parameters are subject to significant degeneracies, particularly among the microphysical parameters ($\epsilon_e$, $\epsilon_B$, and $f_e$), the external density $n_{\rm ext}$, and the kinetic energy ${\mathcal E}_{k}^{\rm iso}$. 
Given the sparse temporal and spectral coverage of the available radio data, multiple combinations of these parameters can reproduce the observed light curves comparably well. 
As a result, the inferred parameter values should be interpreted as representative rather than uniquely determined.

To describe the CNM around the SMBH, we approximate the density profile by a single power-law, $n(R)\propto R^{-w}$. The shock radius in our calculations, $\sim 10^{19}\,{\rm cm}$, for $T \sim 10^9~{\rm s}$, is comparable to the BH sphere of influence, $r_{\rm inf} \sim 1\times10^{19}(M_{\rm BH}/10^7~M_\odot)^{0.6}\,{\rm cm}$ \cite{Generozov:2015wra,Generozov:2016oon}, being within the Bondi radius, $r_{\rm B} \sim 5\times10^{20}\,(M_{\rm BH}/10^7~M_\odot) (T_{\rm eff}/10^4~{\rm K})^{-1} (\gamma/1.67)^{-1} (\mu/0.6)\,{\rm cm}$, where $T_{\rm eff}$, $\gamma$, and $\mu$ are the temperature, adiabatic index, and mean molecular weight of the interstellar medium (ISM) \citep{Bondi1952}. Note that the classical Bondi solution assumes that the gravitational potential is dominated by the SMBH, and is therefore most directly applicable within the sphere of influence. Although more realistic CNM density structures have also been investigated using analytical and numerical models, including effects of stellar winds, angular momentum, and feedback 
(e.g., Refs.~\citep{Quataert:2003ae,Cuadra:2015jda,Generozov:2016oon,Ressler:2018yhi}). our smooth radial profile of the CNM structure is phenomenologically reasonable. In contrast, Ref.~\citep{Matsumoto:2024eja} adopts a much higher ISM temperature, $T_{\rm eff}\sim 10^7~{\rm K}$, which reduces the Bondi radius to $r_{\rm B} \sim 10^{17}\,{\rm cm}$, so that the CNM density transitions to a constant density ISM at $r_{\rm B}$, and the interaction with the ISM profile could play a role in producing the observed radio flare.

We have also examined the dependence of the detectability on the physical parameters. 
For the delayed wind and delayed jet models, which are predominantly in the deceleration phase but remain outside the deep-Newtonian regime over the epochs considered here, the flux in the optical and x-ray bands ($\nu_a<\nu\leq\nu_c$) follows the synchrotron scalings derived from Eqs.~(\ref{nu_m}), (\ref{Fnu_max}), and (\ref{Fnu_total}),
$F_\nu \propto {{\mathcal E}_{k}^{\rm iso}}^{\frac{2(5s+3)-w(s+5)}{4(5-w)}} n_{\rm ext}^{\frac{19-5s}{4(5-w)}} \epsilon_B^{\frac{1+s}{4}}\epsilon_e^{s-1} f_e^{2-s}T^{\frac{3(7-5s)+4w(s-2)}{2(5-w)}}$.
Similarly, in the radio band the flux scales as given in Eq.~(\ref{Fnu_degen}). 
Among the various model parameters, we highlight the CNM density and the isotropic-equivalent kinetic energy, since these quantities influence the dynamical evolution of the outflow.
For example, in the optical and x-ray bands, an order-of-magnitude decrease in the kinetic energy and/or ambient density would typically lead to a substantial reduction in the flux. 
Depending on the adopted parameters, the optical and x-ray emission may fall below the detection thresholds.
On the other hand, at high radio frequencies around $10^{11}\,\mathrm{Hz}$, even allowing for order-of-magnitude variations in the physical parameters (e.g., the kinetic energy or ambient density), the resulting change in the predicted flux is typically limited to factors of a few. The emission at this frequency would then exceed the ALMA sensitivity for the delayed jet model shown in Fig.~\ref{fig:multi}.
Therefore, the models can still be distinguished through high-frequency radio observations. 
Deep observations with facilities such as ALMA are thus expected to discriminate between the delayed wind and jet scenarios.

\section{Summary}
\label{sec:conclusion}
In this paper, using {\textsc AMES}, we have investigated whether radio-detected TDEs can be explained by delayed outflow models. 
For delayed radio flare TDEs, which exhibit radio brightening at late times, we found that emission from an instantaneous wind alone is insufficient to account for the observed light curves. This indicates that a delayed outflow, either a wind or a jet, is required to reproduce the late-time radio emission.
In contrast, the TDE iPTF16fnl, which does not show a delayed radio flare, can be described by the instantaneous wind model without invoking additional components. For some delayed radio flare TDEs, both delayed wind and delayed jet models can explain the observed radio light curves. This highlights the difficulty in inferring outflow properties based solely on radio observations. 
Our results imply that delayed outflows may be energetic if microphysical parameters are similar to those inferred from gamma-ray burst afterglows. Large kinetic energies of ${\mathcal E}_{k}^{\rm iso}\sim10^{51}-10^{52}$~erg are consistent with the gravitational energy budget, and delayed radio-emitting outflows may preferentially originate from TDEs hosting massive BHs with $M_{\rm BH}\sim10^8~M_\odot$.   

We demonstrated that multiwavelength emission can be used for discriminating among different models. For nearby TDEs, x-ray and optical afterglow emission from delayed jets could be detected by \emph{Chandra} and Rubin, respectively. Multiwavelength follow-up observations could be essential for constraining the nature of outflows in radio-emitting TDEs and to distinguish between different emission scenarios. In particular, x-ray and optical monitoring play a critical role in identifying the properties of the radio emitting outflows. Delayed outflows~\cite{Murase:2020lnu,Mukhopadhyay:2023mld} and underlying activities of disks and coronae~\cite{Murase:2019vdl,Hayasaki2019,Murase:2020lnu} would also be important for searching multimessenger signatures from TDEs. 

\section*{Acknowledgements}
We would like to acknowledge 
Kazumi~Kashiyama,
Asuka~Kuwata,
Tatsuya~Matsumoto,
Mainak~Mukhopadhyay,
Ken~Ohsuga,
Rin~Oikawa,
Shuta~J.~Tanaka, 
Seiji~Toshikage,
Ryo~Yamazaki, and
B.~Theodore~Zhang
for useful discussions.
We also thank the anonymous referee for their helpful comments that improved the paper.
The authors are grateful to the Yukawa Institute for Theoretical Physics at Kyoto University. 
This work was completed in part during the YITP long-term workshop YITP-T-26-02, ``Multi-Messenger Astrophysics in the Dynamic Universe.''
This research was partially supported by JSPS KAKENHI, grant numbers Nos.~25KJ0010 (YS) and 20H05852 (KM).
M.B. acknowledges support from the Eberly Research Fellowship at the Pennsylvania State University and the Simons Collaboration on Extreme Electrodynamics of Compact Sources (SCEECS) Postdoctoral Fellowship at the Wisconsin IceCube Particle Astrophysics Center (WIPAC), University of Wisconsin-Madison. 
The work of K.M. is supported by the NSF Grant Nos. AST-2108466, AST-2108467 and AST-2308021.
J.C. also acknowledges the Nevada Center for Astrophysics and NASA award 80NSSC23M0104 for support.

\appendix

\section{Adopted model parameters}
\label{sec:parameter}

In this appendix, we present the adopted model parameters. 
A summary of our results is provided in Table~\ref{tab:parameter}.

\begin{table*}
\centering
\caption{Parameters for several models are listed.}
\label{tab:parameter}
\renewcommand{\arraystretch}{1.2}
\scalebox{1.2}{
\begin{tabular}{lcccccccccc}
\hline
{\bf Instantaneous wind}~ & {$\theta_0$}~[{\rm rad}] & $v_0$ & ${\mathcal E}_{k}^{\rm iso}$~[erg] & $n_{\rm ext}$~${[\rm cm^{-3}]}$ & $w$ & $s$ & $\epsilon_B$ & $\epsilon_e$ & $f_e$ & $T_{\rm del}$~[s]  \\
\hline
ASASSN-14ae${}^{\rm a}$ & \multirow{10}{*}{$-{}^{\rm d}$} & \multirow{10}{*}{$0.1c$} & \multirow{10}{*}{$5\times10^{51}$} & $10^5$ & $1.0$ & $2.8$ & $10^{-3}$ & $0.1$ & $0.1$ & \multirow{10}{*}{$-{}^{\rm e}$}  \\
ASASSN-15oi${}^{\rm a,c}$  & & & & $20$ & $1.5$ & $2.6$ & $8\times10^{-2}$ & $0.4$ & $0.1$ & \\
PS16dtm${}^{\rm a}$  & & & & $10^3$ & $1.0$ & $2.8$ & $10^{-2}$ & $0.1$ & $0.1$ & \\
iPTF16fnl${}^{\rm b}$  & & & & $20$ & $1.0$ & $2.8$ & $6\times10^{-2}$ & $0.3$ & $0.8$ & \\
ASASSN-19bt${}^{\rm a}$  & & & & $20$ & $1.0$ & $2.2$ & $4\times10^{-2}$ & $0.4$ & $0.1$ & \\
AT~2019dsg${}^{\rm a,c}$  & & & & $4\times10^2$ & $1.0$ & $2.6$ & $4\times10^{-2}$ & $0.3$ & $0.1$ & \\
AT~2019ehz${}^{\rm a}$  & & & & $10^2$ & $2.0$ & $2.7$ & $8\times10^{-2}$ & $0.4$ & $0.85$ & \\
AT~2019eve${}^{\rm a}$  & & & & $5\times10^2$ & $1.0$ & $2.2$ & $6\times10^{-2}$ & $0.3$ & $0.8$ & \\
AT~2020vwl${}^{\rm a,c}$  & & & & $10^2$ & $1.0$ & $2.8$ & $3\times10^{-2}$ & $0.3$ & $0.1$ & \\
\hline
{\bf Delayed wind} & {$\theta_0$}~[{\rm rad}] & $v_0$ & ${\mathcal E}_{k}^{\rm iso}$~[erg] & $n_{\rm ext}$~${[\rm cm^{-3}]}$ & $w$ & $s$ & $\epsilon_B$ & $\epsilon_e$ & $f_e$ & $T_{\rm del}$~[s]  \\
\hline 
ASASSN-14ae${}^{\rm a}$ & \multirow{10}{*}{$-{}^{\rm d}$} & \multirow{10}{*}{$0.1c$} & $5\times10^{52}$ & $10^5$ & $1.0$ & $2.8$ & $10^{-3}$ & $0.1$ & $0.1$ & $2.2\times10^8$  \\
ASASSN-15oi${}^{\rm a,c}$ & & & $5\times10^{52}$ & $20$ & $1.5$ & $2.1$ & $8\times10^{-2}$ & $0.4$ & $0.8$ & $6.0\times10^7$ \\
PS16dtm${}^{\rm a}$ & & & $5\times10^{52}$ & $10^3$ & $1.0$ & $2.4$ & $10^{-2}$ & $0.1$ & $0.2$ & $4.0\times10^7$ \\
iPTF16fnl${}^{\rm b}$ & & & ${5\times10^{51}}$ & $20$ & $1.0$ & $2.8$ & $6\times10^{-2}$ & $0.3$ & $0.8$ & $7.0\times10^6$ \\
ASASSN-19bt${}^{\rm a}$ & & & $5\times10^{52}$ & $20$ & $1.0$ & $2.1$ & $4\times10^{-2}$ & $0.4$ & $0.6$ & $10^6$ \\
AT~2019dsg${}^{\rm a,c}$ & & & $5\times10^{52}$ & $4\times10^2$ & $1.0$ & $2.6$ & $10^{-2}$ & $0.1$ & $0.6$ & $8.0\times10^7$ \\
AT~2019ehz${}^{\rm a}$ & & & $5\times10^{52}$ & $10^2$ & $2.0$ & $2.7$ & $10^{-3}$ & $0.1$ & $0.1$ & $10^7$ \\
AT~2019eve${}^{\rm a}$ & & & $5\times10^{52}$ & $5\times10^2$ & $1.0$ & $2.2$ & $10^{-2}$ & $0.1$ & $0.3$ & $10^7$ \\
AT~2020vwl${}^{\rm a,c}$ & & & $5\times10^{52}$ & $10^2$ & $1.0$ & $2.8$ & $2\times10^{-2}$ & $0.1$ & $0.1$ & $5.2\times10^7$ \\
\hline
{\bf Delayed jet} & $\theta_0$~[rad] & $\Gamma_0$ & ${\mathcal E}_{k}^{\rm iso}$~[erg] & $n_{\rm ext}$~${[\rm cm^{-3}]}$ & $w$ & $s$ & $\epsilon_B$ & $\epsilon_e$ & $f_e$ & $T_{\rm del}$~[s]  \\
\hline
ASASSN-14ae${}^{\rm a}$ & \multirow{10}{*}{$0.35$} & \multirow{10}{*}{$3.0$} & \multirow{10}{*}{$5\times10^{52}$} & $10^5$ & $1.0$ & $2.5$ & $10^{-3}$ & $10^{-3}$ & $0.1$ & $2.2\times10^8$  \\
ASASSN-15oi${}^{\rm a}$ & & & & $20$ & $1.5$ & $2.1$ & $4\times10^{-3}$ & $0.3$ & $0.8$ & $6.0\times10^7$ \\
PS16dtm${}^{\rm a}$ & & & & $10^3$ & $1.0$ & $2.7$ & $2\times10^{-3}$ & $0.1$ & $0.1$ & $4.0\times10^7$ \\
iPTF16fnl${}^{\rm b}$ & & & & $20$ & $1.0$ & $2.2$ & $3\times10^{-5}$ & $10^{-3}$ & $0.1$ & $7.0\times10^6$ \\
ASASSN-19bt${}^{\rm a}$ & & & & $20$ & $1.0$ & $2.8$ & $10^{-3}$ & $3\times10^{-2}$ & $0.6$ & $2.0\times10^7$ \\
AT~2019dsg${}^{\rm a,c}$ & & & & $4\times10^2$ & $1.0$ & $2.8$ & $10^{-4}$ & $0.1$ & $0.1$ & $8.0\times10^7$ \\
AT~2019ehz${}^{\rm a}$ & & & & $10^2$ & $1.0$ & $2.7$ & $3\times10^{-3}$ & $0.1$ & $0.1$ & $10^7$ \\
AT~2019eve${}^{\rm a}$ & & & & $5\times10^2$ & $1.0$ & $2.6$ & $3\times10^{-3}$ & $0.1$ & $0.1$ & $10^7$ \\
AT~2020vwl${}^{\rm a,c}$ & & & & $10^2$ & $1.0$ & $2.6$ & $10^{-3}$ & $2\times10^{-2}$ & $0.1$ & $5.2\times10^7$ \\
\hline
\end{tabular}}
\renewcommand{\arraystretch}{1.0}
\begin{tablenotes}
\item {\bf Notes.} \\
${}^{\rm a}$~Delayed radio flare event.\\
${}^{\rm b}$~Radio event without a delayed flare.\\
${}^{\rm c}$~Rebrightening event.\\
${}^{\rm d}$~N/A (consider an isotropic outflow.)\\
${}^{\rm e}$~N/A.\\
\end{tablenotes}
\end{table*}

\bibliography{TDE}

@article{Murase:2019vdl,
    author = "Murase, Kohta and Kimura, Shigeo S. and Meszaros, Peter",
    title = "{Hidden Cores of Active Galactic Nuclei as the Origin of Medium-Energy Neutrinos: Critical Tests with the MeV Gamma-Ray Connection}",
    eprint = "1904.04226",
    archivePrefix = "arXiv",
    primaryClass = "astro-ph.HE",
    doi = "10.1103/PhysRevLett.125.011101",
    journal = "Phys. Rev. Lett.",
    volume = "125",
    number = "1",
    pages = "011101",
    year = "2020"
}

@article{Murase:2023chr,
    author = "Murase, Kohta",
    title = "{Interacting supernovae as high-energy multimessenger transients}",
    eprint = "2312.17239",
    archivePrefix = "arXiv",
    primaryClass = "astro-ph.HE",
    doi = "10.1103/PhysRevD.109.103020",
    journal = "Phys. Rev. D",
    volume = "109",
    number = "10",
    pages = "103020",
    year = "2024"
}

@article{Mukhopadhyay:2023mld,
    author = "Mukhopadhyay, Mainak and Bhattacharya, Mukul and Murase, Kohta",
    title = "{Multimessenger signatures of delayed choked jets in tidal disruption events}",
    eprint = "2309.02275",
    archivePrefix = "arXiv",
    primaryClass = "astro-ph.HE",
    doi = "10.1093/mnras/stae2080",
    journal = "Mon. Not. Roy. Astron. Soc.",
    volume = "534",
    number = "2",
    pages = "1528--1540",
    year = "2024"
}

@ARTICLE{Hills1975,
       author = {{Hills}, J.~G.},
        title = "{Possible power source of Seyfert galaxies and QSOs}",
      journal = {\nat},
         year = 1975,
        month = mar,
       volume = {254},
       number = {5498},
        pages = {295-298},
          doi = {10.1038/254295a0},
       adsurl = {https://ui.adsabs.harvard.edu/abs/1975Natur.254..295H}
}

@ARTICLE{Rees1988,
    author = "Rees, Martin J.",
    title = "{Tidal disruption of stars by black holes of 10 to the 6th-10 to the 8th solar masses in nearby galaxies}",
    doi = "10.1038/333523a0",
    journal = "Nature",
    volume = "333",
    pages = "523--528",
    year = "1988"
}

@ARTICLE{Komossa2015,
    author = "Komossa, S.",
    title = "{Tidal disruption of stars by supermassive black holes: Status of observations}",
    doi = "10.1016/j.jheap.2015.04.006",
    journal = "JHEAp",
    volume = "7",
    pages = "148--157",
    year = "2015"
}

@ARTICLE{MS2016,
    author = "Metzger, Brian D. and Stone, Nicholas C.",
    title = "{A bright year for tidal disruptions}",
    doi = "10.1093/mnras/stw1394",
    journal = "Mon. Not. Roy. Astron. Soc.",
    volume = "461",
    number = "1",
    pages = "948--966",
    year = "2016"
}

@ARTICLE{EK1989,
       author = {{Evans}, Charles R. and {Kochanek}, Christopher S.},
        title = "{The Tidal Disruption of a Star by a Massive Black Hole}",
      journal = {\apjl},
         year = 1989,
        month = nov,
       volume = {346},
        pages = {L13},
          doi = {10.1086/185567},
       adsurl = {https://ui.adsabs.harvard.edu/abs/1989ApJ...346L..13E}
}

@ARTICLE{DeColle2012,
    author = "De Colle, Fabio and others",
    title = "{The dynamics, appearance and demographics of relativistic jets triggered by tidal disruption of stars in quiescent supermassive black holes}",
    doi = "10.1088/0004-637X/760/2/103",
    journal = "Astrophys. J.",
    volume = "760",
    pages = "103",
    year = "2012"
}

@ARTICLE{van_Velzen2021,
    author = "van Velzen, Sjoert and others",
    title = "{Seventeen Tidal Disruption Events from the First Half of ZTF Survey Observations: Entering a New Era of Population Studies}",
    doi = "10.3847/1538-4357/abc258",
    journal = "Astrophys. J.",
    volume = "908",
    number = "1",
    pages = "4",
    year = "2021"
}

@article{Bloom2011,
    author = "Bloom, Joshua S. and others",
    title = "{A relativistic jetted outburst from a massive black hole fed by a tidally disrupted star}",
    doi = "10.1126/science.1207150",
    journal = "Science",
    volume = "333",
    pages = "203",
    year = "2011"
}

@article{Burrows2011,
    author = "Burrows, D. N. and others",
    title = "{Discovery of the Onset of Rapid Accretion by a Dormant Massive Black Hole}",
    doi = "10.1038/nature10374",
    journal = "Nature",
    volume = "476",
    pages = "421",
    year = "2011"
}

@article{Cenko2012,
    author = "Cenko, S. Bradley and others",
    title = "{Swift J2058.4+0516: Discovery of a Possible Second Relativistic Tidal Disruption Flare}",
    doi = "10.1088/0004-637X/753/1/77",
    journal = "Astrophys. J.",
    volume = "753",
    pages = "77",
    year = "2012"
}

@article{Brown2015,
    author = "Brown, G. C. and others",
    title = "{Swift J1112.2\ensuremath{-}8238: a candidate relativistic tidal disruption flare}",
    doi = "10.1093/mnras/stv1520",
    journal = {\mnras},
    volume = "452",
    number = "4",
    pages = "4297--4306",
    year = "2015"
}

@article{Andreoni2022,
    author = "Andreoni, Igor and others",
    title = "{A very luminous jet from the disruption of a star by a massive black hole}",
    doi = "10.1038/s41586-022-05465-8",
    journal = "\nat",
    volume = "612",
    number = "7940",
    pages = "430--434",
    year = "2022",
    note = "[Erratum: Nature 613, E6 (2023)]"
}

@article{Murase:2020lnu,
    author = "Murase, Kohta and others",
    title = "{High-Energy Neutrino and Gamma-Ray Emission from Tidal Disruption Events}",
    doi = "10.3847/1538-4357/abb3c0",
    journal = "Astrophys. J.",
    volume = "902",
    number = "2",
    pages = "108",
    year = "2020"
}

@ARTICLE{Kelley2014,
    author = "Kelley, Luke Zoltan and Tchekhovskoy, Alexander and Narayan, Ramesh",
    title = "{Tidal Disruption and Magnetic Flux Capture: Powering a Jet from a Quiescent Black Hole}",
    doi = "10.1093/mnras/stu2041",
    journal = "Mon. Not. Roy. Astron. Soc.",
    volume = "445",
    number = "4",
    pages = "3919--3938",
    year = "2014"
}

@ARTICLE{Tchek2014,
       author = {{Tchekhovskoy}, Alexander and others},
    title = "{Swift J1644+57 gone MAD: the case for dynamically-important magnetic flux threading the black hole in a jetted tidal disruption event}",
    doi = "10.1093/mnras/stt2085",
    journal = "Mon. Not. Roy. Astron. Soc.",
    volume = "437",
    number = "3",
    pages = "2744--2760",
    year = "2014"
}

@ARTICLE{Mimica2015,
       author = {{Mimica}, P. and others},
    title = "{The radio afterglow of Swift J1644+57 reveals a powerful jet with fast core and slow sheath}",
    doi = "10.1093/mnras/stv825",
    journal = "Mon. Not. Roy. Astron. Soc.",
    volume = "450",
    number = "3",
    pages = "2824--2841",
    year = "2015"
}

@ARTICLE{Giannios2011,
    author = "Giannios, Dimitrios and Metzger, Brian D.",
    title = "{Radio transients from stellar tidal disruption by massive black holes}",
    doi = "10.1111/j.1365-2966.2011.19188.x",
    journal = "Mon. Not. Roy. Astron. Soc.",
    volume = "416",
    pages = "2102",
    year = "2011"
}

@ARTICLE{NG2007,
    author = "Nakar, Ehud and Granot, Jonathan",
    title = "{Smooth Light Curves from a Bumpy Ride: Relativistic Blast Wave Encounters a Density Jump}",
    doi = "10.1111/j.1365-2966.2007.12245.x",
    journal = "Mon. Not. Roy. Astron. Soc.",
    volume = "380",
    pages = "1744--1760",
    year = "2007"
}

@ARTICLE{Alexander2020,
       author = {{Alexander}, Kate D. and others},
    title = "{Radio Properties of Tidal Disruption Events}",
    doi = "10.1007/s11214-020-00702-w",
    journal = "Space Sci. Rev.",
    volume = "216",
    number = "5",
    pages = "81",
    year = "2020"
}

@ARTICLE{Horesh2021,
    author = "Horesh, Assaf and Cenko, S. Bradley and Arcavi, Iair",
    title = "{Delayed Radio Flares from a Tidal Disruption Event}",
    doi = "10.1038/s41550-021-01300-8",
    journal = "Nature Astron.",
    volume = "5",
    number = "5",
    pages = "491--497",
    year = "2021"
}

@article{Matsumoto:2021qqo,
    author = "Matsumoto, Tatsuya and Piran, Tsvi and Krolik, Julian H.",
    title = "{What powers the radio emission in TDE AT2019dsg: A long-lived jet or the disruption itself?}",
    doi = "10.1093/mnras/stac382",
    journal = "Mon. Not. Roy. Astron. Soc.",
    volume = "511",
    number = "4",
    pages = "5085--5092",
    year = "2022"
}

@article{Cendes2021,
	author = {{Cendes}, Y. and others},
        title = "{Radio Observations of an Ordinary Outflow from the Tidal Disruption Event AT2019dsg}",
      journal = {\apj},
         year = 2021,
        month = oct,
       volume = {919},
       number = {2},
          eid = {127},
        pages = {127},
          doi = {10.3847/1538-4357/ac110a},
       adsurl = {https://ui.adsabs.harvard.edu/abs/2021ApJ...919..127C}
}

@ARTICLE{Piran2015,
       author = {{Piran}, Tsvi and others},
    title = "{''Circularization'' vs. Accretion -- What Powers Tidal Disruption Events?}",
    doi = "10.1088/0004-637X/806/2/164",
    journal = "Astrophys. J.",
    volume = "806",
    pages = "164",
    year = "2015"
}

@ARTICLE{Alexander2016,
       author = {{Alexander}, K.~D. and others},
    title = "{Discovery of an outflow from radio observations of the tidal disruption event ASASSN-14li}",
    doi = "10.3847/2041-8205/819/2/L25",
    journal = "Astrophys. J. Lett.",
    volume = "819",
    number = "2",
    pages = "L25",
    year = "2016"
}

@ARTICLE{Horesh2021b,
       author = {{Horesh}, Assaf and others},
    title = "{Are Delayed Radio Flares Common in Tidal Disruption Events? The Case of the TDE iPTF 16fnl}",
    doi = "10.3847/2041-8213/ac25fe",
    journal = "Astrophys. J. Lett.",
    volume = "920",
    number = "1",
    pages = "L5",
    year = "2021"
}

@ARTICLE{Dai2002,
    author = "Dai, Z. G. and Lu, T.",
    title = "{Hydrodynamics of relativistic blast waves in a density-jump medium and their emission signature}",
    doi = "10.1086/339418",
    journal = "Astrophys. J. Lett.",
    volume = "565",
    pages = "L87--L90",
    year = "2002"
}

@article{Hayasaki2023,
    author = "Hayasaki, Kimitake and Yamazaki, Ryo",
    title = "{Disk Wind\textendash{}Driven Expanding Radio-emitting Shell in Tidal Disruption Events}",
    doi = "10.3847/1538-4357/ace35a",
    journal = "Astrophys. J.",
    volume = "954",
    number = "1",
    pages = "5",
    year = "2023"
}

@ARTICLE{Hayasaki2019,
       author = {{Hayasaki}, Kimitake and {Yamazaki}, Ryo},
        title = "{Neutrino Emissions from Tidal Disruption Remnants}",
      journal = {\apj},
         year = 2019,
        month = dec,
       volume = {886},
       number = {2},
          eid = {114},
        pages = {114},
          doi = {10.3847/1538-4357/ab44ca},
       adsurl = {https://ui.adsabs.harvard.edu/abs/2019ApJ...886..114H}
}

@ARTICLE{Cendes2023,
    author = "Cendes, Yvette and others",
    title = "{Ubiquitous Late Radio Emission from Tidal Disruption Events}",
    eprint = "2308.13595",
    archivePrefix = "arXiv",
    primaryClass = "astro-ph.HE",
    doi = "10.3847/1538-4357/ad5541",
    journal = "Astrophys. J.",
    volume = "971",
    number = "2",
    pages = "185",
    year = "2024"
}

@ARTICLE{Cendes2022,
    author = "Cendes, Yvette and others",
    title = "{A Mildly Relativistic Outflow Launched Two Years after Disruption in Tidal Disruption Event AT2018hyz}",
    doi = "10.3847/1538-4357/ac88d0",
    journal = "Astrophys. J.",
    volume = "938",
    number = "1",
    pages = "28",
    year = "2022"
}

@ARTICLE{Sun2015,
       author = {{Sun}, Hui and {Zhang}, Bing and {Li}, Zhuo},
        title = "{Extragalactic High-energy Transients: Event Rate Densities and Luminosity Functions}",
      journal = {\apj},
         year = 2015,
        month = oct,
       volume = {812},
       number = {1},
          eid = {33},
        pages = {33},
          doi = {10.1088/0004-637X/812/1/33},
       adsurl = {https://ui.adsabs.harvard.edu/abs/2015ApJ...812...33S}
}

@ARTICLE{Sfaradi2023,
       author = {{Sfaradi}, Itai and others},
        title = "{An off-axis relativistic jet seen in the long lasting delayed radio flare of the TDE AT 2018hyz}",
      journal = {\mnras},
         year = 2024,
        month = jan,
       volume = {527},
       number = {3},
        pages = {7672-7680},
          doi = {10.1093/mnras/stad3717},
       adsurl = {https://ui.adsabs.harvard.edu/abs/2024MNRAS.527.7672S}
}

@article{Murase:2010fq,
    author = "Murase, Kohta and others",
    title = "{On the Implications of Late Internal Dissipation for Shallow-Decay Afterglow Emission and Associated High-Energy Gamma-Ray Signals}",
    doi = "10.1088/0004-637X/732/2/77",
    journal = "Astrophys. J.",
    volume = "732",
    pages = "77",
    year = "2011"
}

@ARTICLE{Beniamini2023,
    author = "Beniamini, Paz and Piran, Tsvi and Matsumoto, Tatsuya",
    title = "{Swift J1644+57 as an off-axis Jet}",
    doi = "10.1093/mnras/stad1950",
    journal = "Mon. Not. Roy. Astron. Soc.",
    volume = "524",
    number = "1",
    pages = "1386--1395",
    year = "2023"
}

@article{Matsumoto:2022hqp,
    author = "Matsumoto, Tatsuya and Piran, Tsvi",
    title = "{Generalized equipartition method from an arbitrary viewing angle}",
    doi = "10.1093/mnras/stad1269",
    journal = "\mnras",
    volume = "522",
    number = "3",
    pages = "4565--4576",
    year = "2023"
}

@article{Zhang:2023uei,
    author = "Zhang, B. Theodore and Murase, Kohta and Ioka, Kunihito and Zhang, Bing",
    title = "{The origin of very-high-energy gamma-rays from GRB 221009A: Implications for reverse shock proton synchrotron emission}",
    eprint = "2311.13671",
    archivePrefix = "arXiv",
    primaryClass = "astro-ph.HE",
    reportNumber = "YITP-23-151",
    doi = "10.1016/j.jheap.2025.01.007",
    journal = "JHEAp",
    volume = "45",
    pages = "392--408",
    year = "2025"
}

@ARTICLE{Strubbe2009,
    author = "Strubbe, Linda E. and Quataert, Eliot",
    title = "{Optical Flares from the Tidal Disruption of Stars by Massive Black Holes}",
    doi = "10.1111/j.1365-2966.2009.15599.x",
    journal = "Mon. Not. Roy. Astron. Soc.",
    volume = "400",
    pages = "2070",
    year = "2009"
}

@ARTICLE{Lodato2011,
    author = "Lodato, Giuseppe and Rossi, Elena",
    title = "{Multiband lightcurves of tidal disruption events}",
    doi = "10.1111/j.1365-2966.2010.17448.x",
    journal = "Mon. Not. Roy. Astron. Soc.",
    volume = "410",
    pages = "359",
    year = "2011"
}

@ARTICLE{Roth2016,
       author = {{Roth}, Nathaniel and others},
        title = "{The X-Ray through Optical Fluxes and Line Strengths of Tidal Disruption Events}",
      journal = {\apj},
         year = 2016,
        month = aug,
       volume = {827},
       number = {1},
          eid = {3},
        pages = {3},
          doi = {10.3847/0004-637X/827/1/3},
       adsurl = {https://ui.adsabs.harvard.edu/abs/2016ApJ...827....3R}
}

@ARTICLE{Loeb1997,
    author = "Loeb, Abraham and Ulmer, Andrew",
    title = "{Optical appearance of the debris of a star disrupted by a massive black hole}",
    doi = "10.1086/304814",
    journal = "Astrophys. J.",
    volume = "489",
    pages = "573",
    year = "1997"
}

@ARTICLE{Zhang2021,
       author = {{Zhang}, B. Theodore and others},
    title = "{External Inverse-Compton Emission from Low-luminosity Gamma-Ray Bursts: Application to GRB 190829A}",
    doi = "10.3847/1538-4357/ac0cfc",
    journal = "Astrophys. J.",
    volume = "920",
    number = "1",
    pages = "55",
    year = "2021"
}

@ARTICLE{van_Velzen2016,
    author = "van Velzen, Sjoert and others",
    title = "{A radio jet from the optical and x-ray bright stellar tidal disruption flare ASASSN-14li}",
    doi = "10.1126/science.aad1182",
    journal = "Science",
    volume = "351",
    number = "6268",
    pages = "62--65",
    year = "2016"
}

@ARTICLE{Bondi1952,
    author = "Bondi, H.",
    title = "{On spherically symmetrical accretion}",
    doi = "10.1093/mnras/112.2.195",
    journal = "Mon. Not. Roy. Astron. Soc.",
    volume = "112",
    pages = "195",
    year = "1952"
}

@article{Goodwin:2023qcq,
    author = "Goodwin, A. J. and others",
    title = "{A radio-emitting outflow produced by the tidal disruption event AT2020vwl}",
    doi = "10.1093/mnras/stad1258",
    journal = "Mon. Not. Roy. Astron. Soc.",
    volume = "522",
    number = "4",
    pages = "5084--5097",
    year = "2023"
}

@ARTICLE{Yuan2024,
    author = "Yuan, Chengchao and Zhang, B. Theodore and Winter, Walter and Murase, Kohta",
    title = "{Structured Jet Model for Multiwavelength Observations of the Jetted Tidal Disruption Event AT 2022cmc}",
    eprint = "2406.11513",
    archivePrefix = "arXiv",
    primaryClass = "astro-ph.HE",
    doi = "10.3847/1538-4357/ad6c50",
    journal = "Astrophys. J.",
    volume = "974",
    number = "2",
    pages = "162",
    year = "2024"
}

@ARTICLE{Sato2024,
    author = "Sato, Yuri and Murase, Kohta and Bhattacharya, Mukul and Carpio, Jose A. and Mukhopadhyay, Mainak and Zhang, B. Theodore",
    title = "{Two-component off-axis jet model for radio flares of tidal disruption events}",
    eprint = "2404.13326",
    archivePrefix = "arXiv",
    primaryClass = "astro-ph.HE",
    doi = "10.1103/PhysRevD.110.L061307",
    journal = "Phys. Rev. D",
    volume = "110",
    number = "6",
    pages = "L061307",
    year = "2024"
}

@ARTICLE{Anumarlapudi2024,
    author = "Anumarlapudi, Akash and others",
    title = "{Radio Afterglows from Tidal Disruption Events: An Unbiased Sample from ASKAP RACS}",
    eprint = "2407.12097",
    archivePrefix = "arXiv",
    primaryClass = "astro-ph.HE",
    doi = "10.3847/1538-4357/ad64d3",
    journal = "Astrophys. J.",
    volume = "974",
    number = "2",
    pages = "241",
    year = "2024"
}

@ARTICLE{Hajela2024,
    author = "Hajela, A. and others",
    title = "{Eight Years of Light from ASASSN-15oi: Toward Understanding the Late-time Evolution of TDEs}",
    eprint = "2407.19019",
    archivePrefix = "arXiv",
    primaryClass = "astro-ph.HE",
    doi = "10.3847/1538-4357/adb620",
    journal = "Astrophys. J.",
    volume = "983",
    number = "1",
    pages = "29",
    year = "2025"
}

@article{Goodwin:2024cyg,
    author = "Goodwin, A. J. and others",
    title = "{A Second Radio Flare from the Tidal Disruption Event AT2020vwl: A Delayed Outflow Ejection?}",
    eprint = "2410.18665",
    archivePrefix = "arXiv",
    primaryClass = "astro-ph.HE",
    doi = "10.3847/1538-4357/adb0b1",
    journal = "Astrophys. J.",
    volume = "981",
    number = "2",
    pages = "122",
    year = "2025"
}

@article{Horesh:2021szq,
    author = "Horesh, Assaf and Sfaradi, Itai and Fender, Rob and Green, David A. and Williams, David R. A. and Bright, Joe S.",
    title = "{Are Delayed Radio Flares Common in Tidal Disruption Events? The Case of the TDE iPTF 16fnl}",
    eprint = "2109.10921",
    archivePrefix = "arXiv",
    primaryClass = "astro-ph.HE",
    doi = "10.3847/2041-8213/ac25fe",
    journal = "Astrophys. J. Lett.",
    volume = "920",
    number = "1",
    pages = "L5",
    year = "2021"
}

@article{Christy:2024szt,
    author = "Christy, Collin T. and others",
    title = "{The Peculiar Radio Evolution of the Tidal Disruption Event ASASSN-19bt}",
    eprint = "2404.12431",
    archivePrefix = "arXiv",
    primaryClass = "astro-ph.HE",
    doi = "10.3847/1538-4357/ad675b",
    journal = "Astrophys. J.",
    volume = "974",
    number = "1",
    pages = "18",
    year = "2024"
}

@misc{Chandra,
     note = {The official website of Chandra, \url{(https://chandra.harvard.edu}}
}

@misc{Rubin,
     note = {The official website of Rubin, \url{https://www.lsst.org}}
}

@misc{ALMA,
     note = {The official website of ALMA, \url{https://alma-telescope.jp/en/}}
}

@misc{SKA,
     note = {The official website of SKAO, \url{https://www.skao.int/en}}
}

@misc{VLA,
     note = {The official website of VLA, \url{https://public.nrao.edu/telescopes/vla/}}
}

@misc{MeerKAT,
     note = {The official website of MeerKAT, \url{https://www.sarao.ac.za/science/meerkat/about-meerkat/}}
}

@misc{ZTF,
     note = {The official website of ZTF, \url{https://www.ztf.caltech.edu}}
}

@misc{CTA,
     note = {The official website of CTAO, \url{https://www.ctao.org}}
}

@article{Hayasaki:2021ggg,
    author = "Hayasaki, Kimitake and Jonker, Peter G.",
    title = "{On the Origin of Late-time X-Ray Flares in UV/optically Selected Tidal Disruption Events}",
    eprint = "2107.03666",
    archivePrefix = "arXiv",
    primaryClass = "astro-ph.HE",
    doi = "10.3847/1538-4357/ac18c2",
    journal = "Astrophys. J.",
    volume = "921",
    number = "1",
    pages = "20",
    year = "2021"
}

@article{Matsumoto:2024eja,
    author = "Matsumoto, Tatsuya and Piran, Tsvi",
    title = "{Late-time Radio Flares in Tidal Disruption Events}",
    eprint = "2404.15966",
    archivePrefix = "arXiv",
    primaryClass = "astro-ph.HE",
    doi = "10.3847/1538-4357/ad58ba",
    journal = "Astrophys. J.",
    volume = "971",
    number = "1",
    pages = "49",
    year = "2024"
}

@article{Zhuang:2024tos,
    author = "Zhuang, Jialun and Shen, Rong-Feng and Mou, Guobin and Lu, Wenbin",
    title = "{Interaction of an Outflow with Surrounding Gaseous Clouds as the Origin of Late-time Radio Flares in Tidal Disruption Events}",
    eprint = "2406.08012",
    archivePrefix = "arXiv",
    primaryClass = "astro-ph.HE",
    doi = "10.3847/1538-4357/ad9b98",
    journal = "Astrophys. J.",
    volume = "979",
    number = "2",
    pages = "109",
    year = "2025"
}

@article{Guillochon:2015qfa,
    author = "Guillochon, James and Ramirez-Ruiz, Enrico",
    title = "{A Dark Year for Tidal Disruption Events}",
    eprint = "1501.05306",
    archivePrefix = "arXiv",
    primaryClass = "astro-ph.HE",
    doi = "10.1088/0004-637X/809/2/166",
    journal = "Astrophys. J.",
    volume = "809",
    number = "2",
    pages = "166",
    year = "2015"
}

@article{Piran:2015ija,
    author = "Piran, Tsvi and Sadowski, Aleksander and Tchekhovskoy, Alexander",
    title = "{Jet and disc luminosities in tidal disruption events}",
    eprint = "1501.02015",
    archivePrefix = "arXiv",
    primaryClass = "astro-ph.HE",
    doi = "10.1093/mnras/stv1547",
    journal = "Mon. Not. Roy. Astron. Soc.",
    volume = "453",
    number = "1",
    pages = "157--165",
    year = "2015"
}

@article{vanVelzen:2020cwu,
    author = "van Velzen, Sjoert and others",
    title = "{Seventeen Tidal Disruption Events from the First Half of ZTF Survey Observations: Entering a New Era of Population Studies}",
    eprint = "2001.01409",
    archivePrefix = "arXiv",
    primaryClass = "astro-ph.HE",
    doi = "10.3847/1538-4357/abc258",
    journal = "Astrophys. J.",
    volume = "908",
    number = "1",
    pages = "4",
    year = "2021"
}

@article{Alexander:2025wnh,
    author = "Alexander, Kate D. and others",
    title = "{The Multiwavelength Context of Delayed Radio Emission in Tidal Disruption Events: Evidence for Accretion-driven Outflows}",
    eprint = "2506.12729",
    archivePrefix = "arXiv",
    primaryClass = "astro-ph.HE",
    doi = "10.3847/1538-4357/ae40ab",
    journal = "Astrophys. J.",
    volume = "1000",
    number = "1",
    pages = "139",
    year = "2026"
}

@article{Wijers:1997xu,
    author = "Wijers, Ralph A. M. J. and Rees, Martin J. and Meszaros, Peter",
    title = "{Shocked by GRB-970228: The Afterglow of a cosmological fireball}",
    eprint = "astro-ph/9704153",
    archivePrefix = "arXiv",
    doi = "10.1093/mnras/288.4.L51",
    journal = "Mon. Not. Roy. Astron. Soc.",
    volume = "288",
    pages = "L51--L56",
    year = "1997"
}

@article{Dai:1999fc,
    author = "Dai, Z. G. and Lu, T.",
    title = "{The afterglow of grb 990123 and a dense medium}",
    eprint = "astro-ph/9904025",
    archivePrefix = "arXiv",
    doi = "10.1086/312127",
    journal = "Astrophys. J. Lett.",
    volume = "519",
    pages = "L155--L158",
    year = "1999"
}

@article{Huang:2003nw,
    author = "Huang, Y. F. and Cheng, K. S.",
    title = "{Gamma-ray bursts: Optical afterglows in the deep Newtonian phase}",
    eprint = "astro-ph/0301387",
    archivePrefix = "arXiv",
    doi = "10.1046/j.1365-8711.2003.06430.x",
    journal = "Mon. Not. Roy. Astron. Soc.",
    volume = "341",
    pages = "263--269",
    year = "2003"
}

@article{Piran:2004ba,
    author = "Piran, Tsvi",
    title = "{The physics of gamma-ray bursts}",
    eprint = "astro-ph/0405503",
    archivePrefix = "arXiv",
    doi = "10.1103/RevModPhys.76.1143",
    journal = "Rev. Mod. Phys.",
    volume = "76",
    pages = "1143--1210",
    year = "2004"
}

@BOOK{Bing2018,
       author = {{Zhang}, Bing},
        title = "{The Physics of Gamma-Ray Bursts}",
         year = 2018,
          doi = {10.1017/9781139226530},
       adsurl = {https://ui.adsabs.harvard.edu/abs/2018pgrb.book.....Z}
}

@ARTICLE{Lin2025,
    author = "Lin, Zikun and Wang, Yanan and Bu, De-Fu and Mao, Junjie and Liu, Jifeng",
    title = "{Delayed Launch of Ultrafast Outflows in the Tidal Disruption Event AT2020afhd}",
    eprint = "2507.15482",
    archivePrefix = "arXiv",
    primaryClass = "astro-ph.HE",
    doi = "10.3847/2041-8213/adf205",
    journal = "Astrophys. J. Lett.",
    volume = "989",
    number = "1",
    pages = "L9",
    year = "2025"
}

@ARTICLE{Sfaradi2025,
    author = "Sfaradi, Itai and others",
    title = "{The First Radio-bright Off-nuclear Tidal Disruption Event AT 2024tvd Reveals the Fastest-evolving Double-peaked Radio Emission}",
    eprint = "2508.03807",
    archivePrefix = "arXiv",
    primaryClass = "astro-ph.HE",
    doi = "10.3847/2041-8213/ae0a26",
    journal = "Astrophys. J. Lett.",
    volume = "992",
    number = "2",
    pages = "L18",
    year = "2025"
}

@ARTICLE{Golay2025,
    author = "Golay, Walter W. and others",
    title = "{Radio Emission from the Infrared Tidal Disruption Event WTP14adeqka: A Directly Resolved Nonrelativistic Outflow}",
    eprint = "2508.16756",
    archivePrefix = "arXiv",
    primaryClass = "astro-ph.HE",
    doi = "10.3847/1538-4357/ae8536",
    journal = "Astrophys. J.",
    volume = "1007",
    number = "1",
    pages = "88",
    year = "2026"
}

@article{Metzger:2022lob,
    author = "Metzger, Brian D.",
    title = "{Cooling Envelope Model for Tidal Disruption Events}",
    eprint = "2207.07136",
    archivePrefix = "arXiv",
    primaryClass = "astro-ph.HE",
    doi = "10.3847/2041-8213/ac90ba",
    journal = "Astrophys. J. Lett.",
    volume = "937",
    number = "1",
    pages = "L12",
    year = "2022"
}

@ARTICLE{Wu2025,
    author = "Wu, Samantha C. and Tsuna, Daichi and Mockler, Brenna and Piro, Anthony L.",
    title = "{Delayed Radio Emission in Tidal Disruption Events from Collisions of Outflows Driven by Disk Instabilities}",
    eprint = "2511.14008",
    archivePrefix = "arXiv",
    primaryClass = "astro-ph.HE",
    doi = "10.3847/1538-4357/ae36a3",
    journal = "Astrophys. J.",
    volume = "998",
    number = "2",
    pages = "199",
    year = "2026"
}

@article{Yuan:2021jjt,
    author = "Yuan, Chengchao and Murase, Kohta and Zhang, B. Theodore and Kimura, Shigeo S. and M{\'e}sz{\'a}ros, Peter",
    title = "{Post-Merger Jets from Supermassive Black Hole Coalescences as Electromagnetic Counterparts of Gravitational Wave Emission}",
    eprint = "2101.05788",
    archivePrefix = "arXiv",
    primaryClass = "astro-ph.HE",
    doi = "10.3847/2041-8213/abee24",
    journal = "Astrophys. J. Lett.",
    volume = "911",
    number = "1",
    pages = "L15",
    year = "2021"
}

@article{Yuan:2024sxk,
    author = "Yuan, Chengchao and Winter, Walter and Zhang, B. Theodore and Murase, Kohta and Zhang, Bing",
    title = "{Revisiting X-Ray Afterglows of Jetted Tidal Disruption Events with the External Reverse Shock}",
    eprint = "2411.07925",
    archivePrefix = "arXiv",
    primaryClass = "astro-ph.HE",
    doi = "10.3847/1538-4357/adbbde",
    journal = "Astrophys. J.",
    volume = "982",
    number = "2",
    pages = "196",
    year = "2025"
}

@ARTICLE{Miller2025,
       author = {{Miller}, J.~M. and {Mockler}, B.},
        title = "{Archival Swift Observations of the Host Galaxy of the TDE AT2025aarm}",
      journal = {The Astronomer's Telegram},
         year = 2025,
        month = oct,
       volume = {17466},
        pages = {1},
       adsurl = {https://ui.adsabs.harvard.edu/abs/2025ATel17466....1M}
}

@ARTICLE{CTAO2025,
       author = {{Paneque}, D. and {Teshima}, M. and {Kherlakian}, M. and {Simongini}, A. and {Katagiri}, H. and {Schiavone}, F. and {Menon}, S. and {Inoue}, S. and {Martins}, V.~B.},
        title = "{Gamma-ray upper limits the on the rising phase of TDE 2025aarm from observations by the MAGIC and LST-1 telescopes}",
      journal = {Transient Name Server AstroNote},
         year = 2025,
        month = nov,
       volume = {328},
        pages = {1},
       adsurl = {https://ui.adsabs.harvard.edu/abs/2025TNSAN.328....1P}
}

@article{Teboul:2023bpk,
    author = "Teboul, Odelia and Metzger, Brian D.",
    title = "{A Unified Theory of Jetted Tidal Disruption Events: From Promptly Escaping Relativistic to Delayed Transrelativistic Jets}",
    eprint = "2308.05161",
    archivePrefix = "arXiv",
    primaryClass = "astro-ph.HE",
    doi = "10.3847/2041-8213/ad0037",
    journal = "Astrophys. J. Lett.",
    volume = "957",
    number = "1",
    pages = "L9",
    year = "2023"
}

@article{Krolik:2024tvu,
    author = "Krolik, Julian and Piran, Tsvi and Ryu, Taeho",
    title = "{Follow the Mass{\textemdash}A Concordance Picture of Tidal Disruption Events}",
    eprint = "2409.02894",
    archivePrefix = "arXiv",
    primaryClass = "astro-ph.HE",
    doi = "10.3847/1538-4357/ade797",
    journal = "Astrophys. J.",
    volume = "988",
    number = "2",
    pages = "220",
    year = "2025"
}

@ARTICLE{Mummery2025,
       author = {{Mummery}, Andrew and {Metzger}, Brian and {van Velzen}, Sjoert and {Guolo}, Muryel},
        title = "{Tidal disruption event Calorimetry: Observational constraints on the physics of TDE optical flares}",
      journal = {arXiv e-prints},
         year = 2025,
        month = dec,
          eid = {arXiv:2512.09143},
        pages = {arXiv:2512.09143},
          doi = {10.48550/arXiv.2512.09143},
archivePrefix = {arXiv},
       eprint = {2512.09143},
 primaryClass = {astro-ph.HE},
       adsurl = {https://ui.adsabs.harvard.edu/abs/2025arXiv251209143M}
}

@ARTICLE{Leventis2012,
       author = {{Leventis}, K. and {van Eerten}, H.~J. and {Meliani}, Z. and {Wijers}, R.~A.~M.~J.},
        title = "{Practical flux prescriptions for gamma-ray burst afterglows, from early to late times}",
      journal = {\mnras},
         year = 2012,
        month = dec,
       volume = {427},
       number = {2},
        pages = {1329-1343},
          doi = {10.1111/j.1365-2966.2012.21994.x},
archivePrefix = {arXiv},
       eprint = {1206.2848},
 primaryClass = {astro-ph.HE},
       adsurl = {https://ui.adsabs.harvard.edu/abs/2012MNRAS.427.1329L}
}

@article{Quataert:2003ae,
    author = "Quataert, Eliot",
    title = "{A dynamical model for hot gas in the galactic center}",
    eprint = "astro-ph/0310446",
    archivePrefix = "arXiv",
    doi = "10.1086/422973",
    journal = "Astrophys. J.",
    volume = "613",
    pages = "322--325",
    year = "2004"
}

@article{Cuadra:2015jda,
    author = "Cuadra, Jorge and Nayakshin, Sergei and Wang, Q. Daniel",
    title = "{The role of feedback in accretion on Low Luminosity AGN: Sgr A* case study}",
    eprint = "1503.02745",
    archivePrefix = "arXiv",
    primaryClass = "astro-ph.HE",
    doi = "10.1093/mnras/stv584",
    journal = "Mon. Not. Roy. Astron. Soc.",
    volume = "450",
    number = "1",
    pages = "277--287",
    year = "2015"
}

@article{Generozov:2016oon,
    author = "Generozov, A. and Mimica, P. and Metzger, B. D. and Stone, N. C. and Giannios, D. and Aloy, M. A.",
    title = "{The influence of circumnuclear environment on the radio emission from TDE jets}",
    eprint = "1605.08437",
    archivePrefix = "arXiv",
    primaryClass = "astro-ph.HE",
    doi = "10.1093/mnras/stw2439",
    journal = "Mon. Not. Roy. Astron. Soc.",
    volume = "464",
    number = "2",
    pages = "2481--2498",
    year = "2017"
}

@article{Ressler:2018yhi,
    author = "Ressler, Sean M. and Quataert, Eliot and Stone, James M.",
    title = "{Hydrodynamic simulations of the inner accretion flow of Sagittarius A* fuelled by stellar winds}",
    eprint = "1805.00474",
    archivePrefix = "arXiv",
    primaryClass = "astro-ph.HE",
    doi = "10.1093/mnras/sty1146",
    journal = "Mon. Not. Roy. Astron. Soc.",
    volume = "478",
    number = "3",
    pages = "3544--3563",
    year = "2018"
}

@article{Generozov:2015wra,
    author = "Generozov, Aleksey and Stone, Nicholas C. and Metzger, Brian D.",
    title = "{Circumnuclear Media of Quiescent Supermassive Black Holes}",
    eprint = "1505.00268",
    archivePrefix = "arXiv",
    primaryClass = "astro-ph.HE",
    doi = "10.1093/mnras/stv1607",
    journal = "Mon. Not. Roy. Astron. Soc.",
    volume = "453",
    number = "1",
    pages = "775--796",
    year = "2015"
}

@ARTICLE{Goodwin2026,
       author = {{Goodwin}, Adelle J. and {Mummery}, Andrew},
        title = "{A universal critical accretion rate for black hole jet formation}",
      journal = {arXiv e-prints},
         year = 2026,
        month = feb,
          eid = {arXiv:2602.14838},
        pages = {arXiv:2602.14838},
          doi = {10.48550/arXiv.2602.14838},
archivePrefix = {arXiv},
       eprint = {2602.14838},
 primaryClass = {astro-ph.HE},
       adsurl = {https://ui.adsabs.harvard.edu/abs/2026arXiv260214838G}
}

@article{Gezari:2021bmb,
    author = "Gezari, Suvi",
    title = "{Tidal Disruption Events}",
    eprint = "2104.14580",
    archivePrefix = "arXiv",
    primaryClass = "astro-ph.HE",
    doi = "10.1146/annurev-astro-111720-030029",
    journal = "Ann. Rev. Astron. Astrophys.",
    volume = "59",
    pages = "21--58",
    year = "2021"
}

@ARTICLE{MacLeod2012,
       author = {{MacLeod}, Morgan and {Guillochon}, James and {Ramirez-Ruiz}, Enrico},
        title = "{The Tidal Disruption of Giant Stars and their Contribution to the Flaring Supermassive Black Hole Population}",
      journal = {\apj},
         year = 2012,
        month = oct,
       volume = {757},
       number = {2},
          eid = {134},
        pages = {134},
          doi = {10.1088/0004-637X/757/2/134},
archivePrefix = {arXiv},
       eprint = {1206.2922},
 primaryClass = {astro-ph.HE},
       adsurl = {https://ui.adsabs.harvard.edu/abs/2012ApJ...757..134M}
}

@article{Hayasaki:2015pxa,
    author = "Hayasaki, Kimitake and Stone, Nicholas C. and Loeb, Abraham",
    title = "{Circularization of Tidally Disrupted Stars around Spinning Supermassive Black Holes}",
    eprint = "1501.05207",
    archivePrefix = "arXiv",
    primaryClass = "astro-ph.HE",
    doi = "10.1093/mnras/stw1387",
    journal = "Mon. Not. Roy. Astron. Soc.",
    volume = "461",
    number = "4",
    pages = "3760--3780",
    year = "2016"
}

@article{Bonnerot:2015ara,
    author = "Bonnerot, Cl{\'e}ment and Rossi, Elena M. and Lodato, Giuseppe and Price, Daniel J.",
    title = "{Disc formation from tidal disruptions of stars on eccentric orbits by Schwarzschild black holes}",
    eprint = "1501.04635",
    archivePrefix = "arXiv",
    primaryClass = "astro-ph.HE",
    doi = "10.1093/mnras/stv2411",
    journal = "Mon. Not. Roy. Astron. Soc.",
    volume = "455",
    number = "2",
    pages = "2253--2266",
    year = "2016"
}

@misc{Gaurava2026,
title = "X-ray Brightening of the Tidal Disruption Event AT 2025aarm",
author = "Jaisawal, \{Gaurava Kumar\} and Hannah Wichern and Panos Charalampopoulos",
year = "2026",
month = apr,
day = "14",
series = "AstroNote",
number = "2026-101",
type = "Other",
}

@ARTICLE{Lu2016,
       author = {{Lu}, Wenbin and {Kumar}, Pawan},
        title = "{External inverse-Compton emission from jetted tidal disruption events}",
      journal = {\mnras},
         year = 2016,
        month = may,
       volume = {458},
       number = {1},
        pages = {1071-1082},
          doi = {10.1093/mnras/stv2324},
archivePrefix = {arXiv},
       eprint = {1602.01468},
 primaryClass = {astro-ph.HE},
       adsurl = {https://ui.adsabs.harvard.edu/abs/2016MNRAS.458.1071L}
}

@ARTICLE{Minhajur2026,
       author = {{Minhajur Rahaman}, Sk. and {Granot}, Jonathan and {Beniamini}, Paz},
        title = "{The Deep Newtonian Regime in Late-Time Blast Waves: Inevitable Transition and Distinct Flux Signatures}",
      journal = {arXiv e-prints},
         year = 2026,
        month = apr,
          eid = {arXiv:2604.23567},
        pages = {arXiv:2604.23567},
          doi = {10.48550/arXiv.2604.23567},
archivePrefix = {arXiv},
       eprint = {2604.23567},
 primaryClass = {astro-ph.HE},
       adsurl = {https://ui.adsabs.harvard.edu/abs/2026arXiv260423567M}
}

@article{Sironi:2013tva,
    author = "Sironi, Lorenzo and Giannios, Dimitrios",
    title = "{A Late-Time Flattening of Light Curves in Gamma-Ray Burst Afterglows}",
    eprint = "1307.3250",
    archivePrefix = "arXiv",
    primaryClass = "astro-ph.HE",
    doi = "10.1088/0004-637X/778/2/107",
    journal = "Astrophys. J.",
    volume = "778",
    pages = "107",
    year = "2013"
}

\end{document}